\documentclass[twocolumn]{aastex631}

\UseRawInputEncoding

\newcommand{\rrab}{RR{\sl ab}}
\newcommand{\rrc}{RR{\sl c}}
\newcommand{\rrd}{RR{\sl d}}

\shorttitle{Variable Stars in Antlia 2}
\shortauthors{Vivas et al.}

\begin{document}

\title{Variable Stars in the giant satellite galaxy Antlia 2}

\author[0000-0003-4341-6172]{A.~Katherina~Vivas}
\affiliation{Cerro Tololo Inter-American Observatory/NSF’s NOIRLab, Casilla 603, La Serena, Chile}

\author[0000-0002-9144-7726]{Clara~E.~Mart{\'\i}nez-V\'azquez}
\affiliation{Cerro Tololo Inter-American Observatory/NSF’s NOIRLab, Casilla 603, La Serena, Chile}
\affiliation{Gemini Observatory/NSFʼs NOIRLab, 670 N. A’ohoku Place, Hilo, HI 96720, USA}

\author[0000-0002-7123-8943]{Alistair~R.~Walker}
\affiliation{Cerro Tololo Inter-American Observatory/NSF’s NOIRLab, Casilla 603, La Serena, Chile}

\author[0000-0002-0038-9584]{Vasily~Belokurov}
\affiliation{Institute of Astronomy, University of Cambridge, Madingley Road, Cambridge CB3 0HA, UK}

\author[0000-0002-9110-6163]{Ting~S.~Li}
\affiliation{Department of Astronomy and Astrophysics, University of Toronto, 50 St. George Street, Toronto ON, M5S 3H4, Canada}

\author[0000-0002-8448-5505]{Denis~Erkal}
\affiliation{Department of Physics, University of Surrey, Guildford GU2 7XH, UK}

\begin{abstract}

We report 350 pulsating variable stars found in four DECam fields ($\sim 12$ sq. deg.) covering the Antlia 2 satellite galaxy. The sample of variables includes 318 RR Lyrae stars and eight anomalous Cepheids in the galaxy. Reclassification of several objects designated previously to be RR Lyrae as Anomalous Cepheids gets rid of the satellite's stars intervening along the line of sight. This in turn removes the need for prolific tidal disruption of the dwarf, in agreement with the recently updated proper motion and peri-centre measurements based on {\it Gaia} EDR3. There are also several bright foreground RR Lyrae stars in the field, and two distant background variables located $\sim 45$ kpc behind Antlia 2. We found RR Lyrae stars over the full search area, suggesting that the galaxy is very large and likely extends beyond our observed area. The mean period of the \rrab{} in Antlia 2 is 0.599 days, while the \rrc{} have a mean period of 0.368 days, indicating the galaxy is an Oosterhoff-intermediate system.  The distance to Antlia 2 based on the RR Lyrae stars is $124.1$ kpc ($\mu_0=20.47$) with a dispersion of $5.4$ kpc. We measured a clear distance gradient along the semi-major axis of the galaxy, with the South-East side of Antlia 2 being $\sim13$ kpc farther away from the North-West side. This elongation along the line of sight is likely due to the ongoing tidal disruption of Ant 2.  

\end{abstract}

\section{Introduction} \label{sec:intro}

The environment of our Milky Way Galaxy (MW) contains more than 50 known smaller galaxies, that range in size from large dwarf galaxies such as the Magellanic Clouds with active star formation and absolute magnitude $M_V \sim -18$, to the many quiescent ultra faint dwarf (UFD) galaxies with luminosities similar to globular clusters \citep{simon19}, the faintest of which have $M_V \sim 0$.  Discovered  mostly by deep wide-field imaging surveys in the past two decades \citep[see reviews by][]{Willman10,mcconnachie12,Belokurov13,simon19} many of these low stellar mass galaxies show signs of their interaction with the MW by elongated shape or by associated tidal debris. The plethora of stellar streams that are now known \citep[see e.g.][]{shipp18} indicates that disruption events may be common.  Analyses of the structure of our galaxy, together with chemo-dynamic analyses of the stellar content of the thick disk and halo, show that encounters and mergers have had a substantial effect on the properties of the MW over its lifetime,  and are continuing to do so. 

\begin{figure}
\plotone{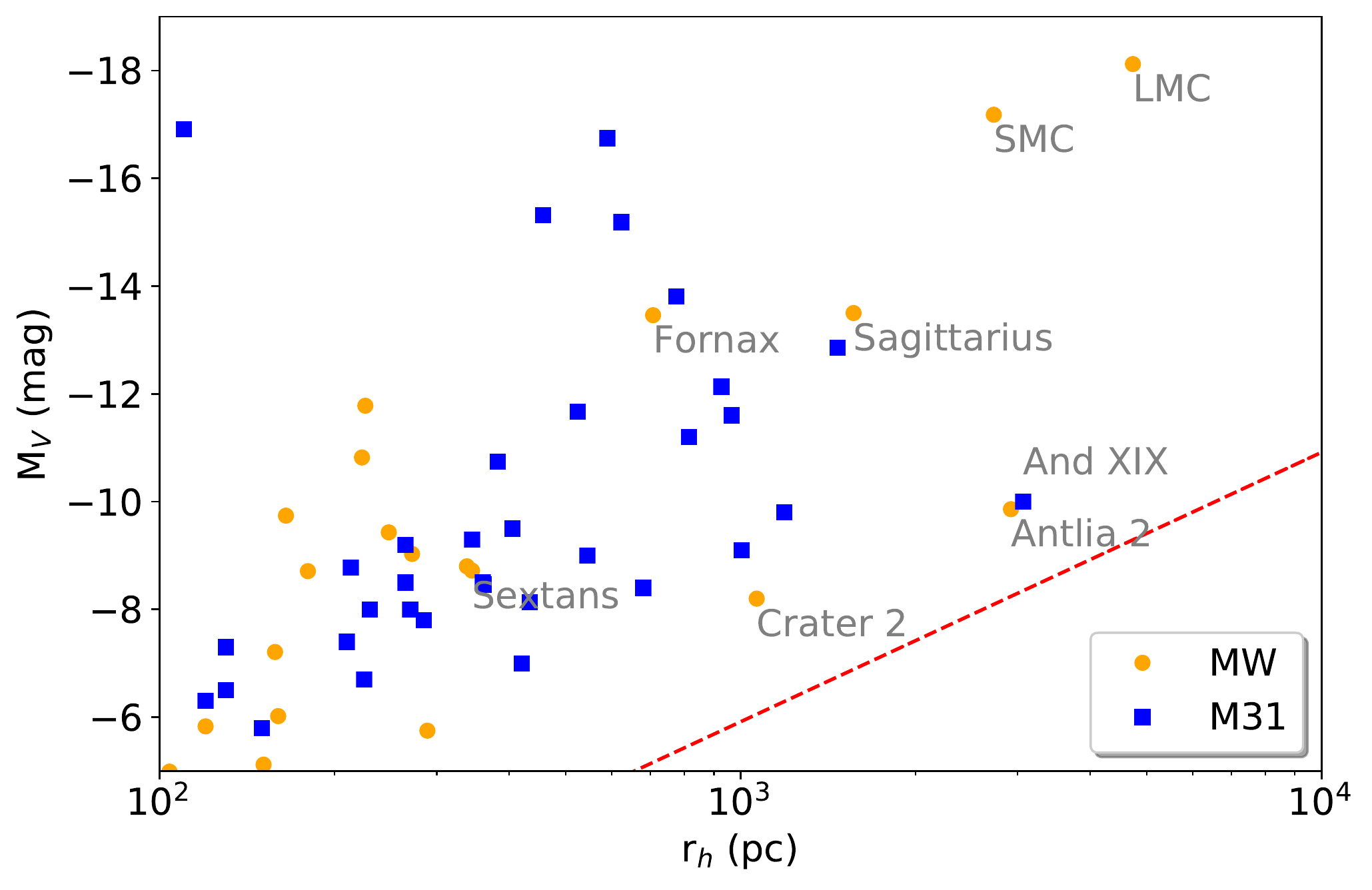}
\caption{Absolute magnitude versus half-light radius for satellite galaxies of the MW and M31 \citep{mcconnachie12,martin16,drlica20,Ji21}. This plot zooms into the region of the brightest and largest satellite galaxies, with most UFDs lying outside the plotted region. The red dotted line indicates surface brightness $\mu=32$ mag/arcsec$^2$.}
\label{fig:Mv-rh}
\end{figure}

A convenient way to view the totality of the MW companion galaxies is by the size-luminosity plot \citep[see for example Figure 11 in][]{T19} which is well-populated over 18 magnitudes in luminosity and half light radii ($r_h$) from 20 to $\sim$4000 pc. However, the completeness as a function of position in this diagram is not well-known except for the brighter systems. Apart from galaxies hidden behind the Galactic plane, there is possible confusion between low luminosity globular clusters and the fainter UFDs, with the compilation of the fainter galaxies also likely to be severely incomplete.  Future surveys such as the Vera C. Rubin Legacy Survey of Space and Time (LSST) are expected to help redress this situation. Figure~\ref{fig:Mv-rh} shows a zoom of the size-luminosity plot in the region of the brightest and largest of the satellite galaxies. The low luminosity/large size region in Figure~\ref{fig:Mv-rh} appears uninhabited, but this could be an observational effect as such low surface brightness galaxies will be dominated by Galactic foreground stars.  Indeed, the discovery of Crater 2 by \citet{torrealba16}  and, in particular, the subsequent discovery of Antlia 2 (Ant 2) by \citet{T19} from \textit{Gaia} proper motions in a field with much higher foreground contamination than Crater 2, shows how nearly invisible such objects can be\footnote{these galaxies can also be found under the names Antlia II and Crater II in the literature}.

\begin{figure*}
\includegraphics[width=0.95\textwidth]{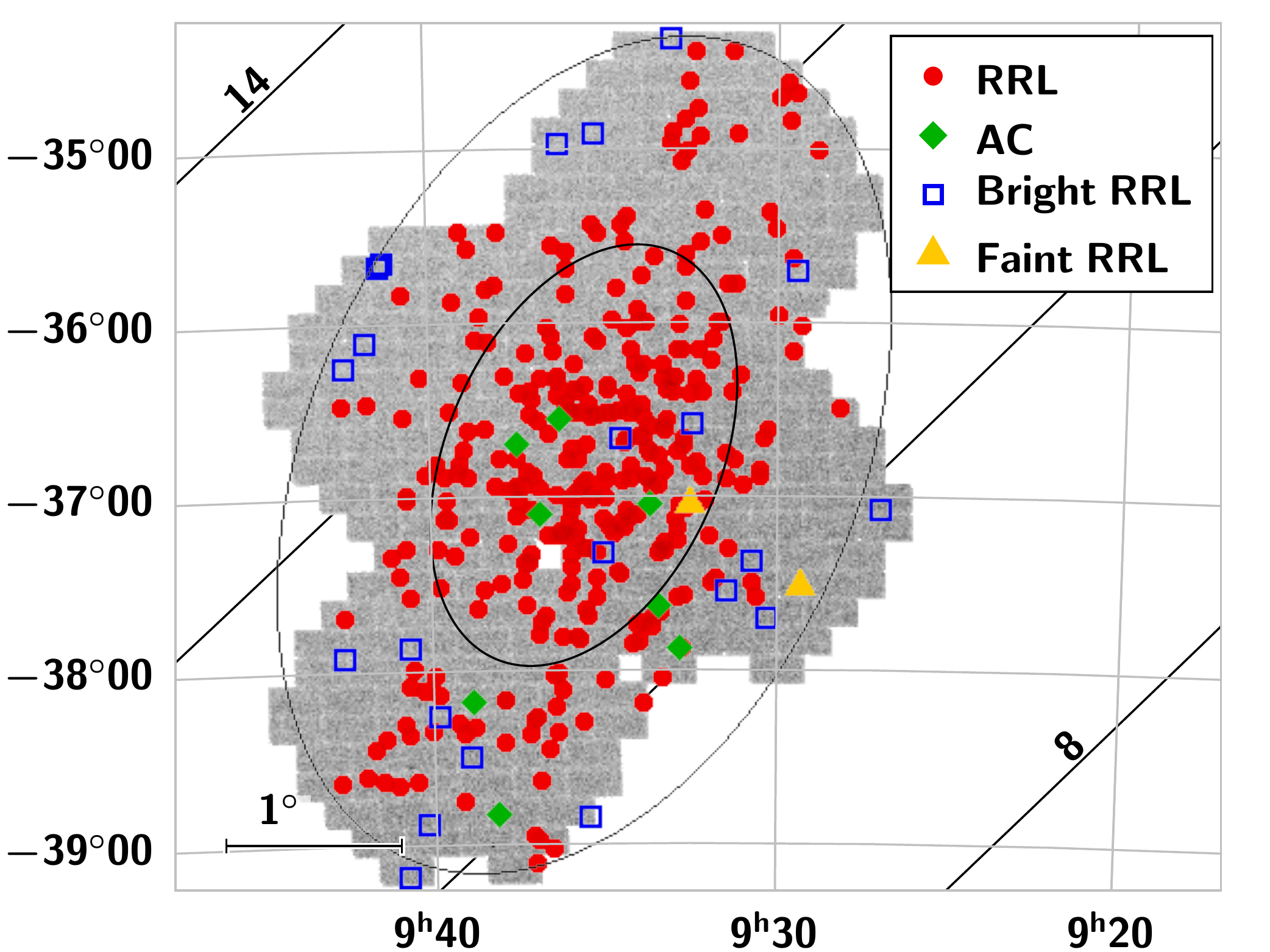}
\caption{Spatial coverage of our DECam observations of Ant 2 in equatorial coordinates. The grey background, which shows the footprint of our survey, is a density map made with all stars in our catalog. The two black ellipses indicate the $1\, r_h$ and $2\, r_h$, as measured by \citet{T19}. Colored symbols indicate the location of pulsating variable stars found in the field: red circles and green diamonds are RRL stars and AC in Ant 2, while open blue squares indicate the location of bright RRL stars which are presumably field halo stars. The two yellow triangles show two faint RRL stars located beyond Ant 2. For reference the diagonal lines indicate lines of constant galactic latitude.}
\label{fig:Sky}
\end{figure*}

The RR Lyrae (RRL) variable stars are a powerful tool for studying the morphology, metallicity and age of, in particular, low surface brightness galaxies \citep{catelan15}. Ubiquitous in any population more than $\sim 10$ Gyr old \citep{savino20}, they are easy to identify even in very crowded fields, and being standard candles they can be definitively assigned (or not) to a minority population galaxy immersed in the MW.  Apart from the rare case of Leo T \citep[][]{clementini12}, all other UFD galaxies so far known terminated their star formation early \citep[$\sim 10$ Gyr or older][]{simon19} and so the RRL stars are representative of the dominant (or only) population present in these galaxies.  These properties of RRL stars are particularly useful when studying galaxies like Ant 2 which, being both a low surface brightness object and located at a galactic latitude of $b\sim +11\degr$, suffers from a very high foreground contamination. 

The discovery of Ant 2 by \citet{T19} was triggered by the identification of a group of three RRL stars in the Gaia DR2 catalog \citep{clementini19} which were close together in the sky, shared similar proper motions, and were located at roughly the same distance. Further investigation by the discovery team however concluded that the RRL stars were not actually part of Ant 2 since they were located in front of it. \citet{T19} suggested that those three RRL stars were the near side of a cloud of debris material originated from Ant 2 during its disruption by the tidal forces of the Milky Way. A complete survey of RRL stars in this galaxy is needed to confirm/deny such a scenario.

In this paper we describe the search and discovery of RRL stars and other variables, either in Ant 2 or its line of sight, from observations made with the Dark Energy Camera \citep[DECam,][]{flaugher15}, and discuss the properties of Ant 2 from an analysis of their pulsational properties and distribution.  In \S~\ref{sec:observations} we describe the observations, data reduction and time series analysis. The population of RRL stars in Ant 2 is discussed in detail in \S~\ref{sec:rrl}, and we  compared their spatial distribution with a model of Ant 2's tidal disruption in \S~\ref{sec:model}. Anomalous Cepheids in Ant 2 are described in \S~\ref{sec:AC}. Finally, conclusions are presented in \S~\ref{sec:conclusion}, with time series data tabulated in Appendix \S~\ref{sec:data}.

\section{Observations} \label{sec:observations}

The Ant 2 observations presented here were collected using DECam on the V{\'\i}ctor M. Blanco 4-meter telescope at Cerro Tololo Inter-American Observatory (CTIO), Chile, on two part nights in November and December 2018, and two full nights in January 2019, as detailed in Table~\ref{tab:Table1}. All these nights were close to Full Moon, therefore we decided to observe in the $r$ and $i$ filters, with exposures 180s and 240s respectively. With its 3 sq. deg. field of view (FoV), DECam is an ideal instrument to study this large galaxy \citep[$r_h=1\fdg 27$,][]{T19}. We used four pointings (Table~\ref{tab:coo}), almost non-overlapping, to cover the galaxy (see Figure~\ref{fig:Sky}). During photometric conditions standard star fields were also observed. 

\begin{table}
	\centering
	\caption{Log of the DECam Ant 2 observations.}
	\label{tab:Table1}
	\begin{tabular}{lccr} 
		\hline
		Run date & $r,i$     & Image Quality & Comments\\
		         & exposures & (arcsec)      &         \\
		\hline
		2018 Nov 23 & 32 & 0.8 & clear\\
		2018 Dec 19 & 56 & 1.0 & clear\\
		2019 Jan 20 & 71 & 1.0 & thin cirrus\\
		2019 Jan 21 & 117 & 1.0 & thin cirrus \\
		\hline
	\end{tabular}
\end{table}

The observational strategy consisted of imaging consecutively the four fields in $r$ and $i$, and then repeating the sequence but with an offset of $75\arcsec$ and $60\arcsec$ in $\alpha$ and $\delta$, respectively,  to fill in the gaps between CCDs. During a night the typical time between observations of the same field/filter was $\sim 30$ min. 

\begin{table}
	\centering
	\caption{Central coordinates of observed fields}
	\label{tab:coo}
	\begin{tabular}{lcc} 
		\hline
		Field & $\alpha$ (J2000.0) & $\delta$ (J2000.0) \\
		      &        (deg)       &      (deg)         \\
		\hline
		A & 143.08400 & -35.32892 \\
		B & 144.81242 & -36.43172 \\
		C & 142.87454 & -37.10664 \\
		D & 144.81225 & -38.25325 \\
		\hline
	\end{tabular}
\end{table}

\subsection{Data Reduction and Photometry} \label{sec:phot}

The images were processed by the DECam Community Pipeline \citep{valdes14}, which removes the instrument signature and provides several data products\footnote{Raw and reduced images are publicly available through the NOIRLab Astro Data Archive (\url{https://astroarchive.noirlab.edu}), under propID 2018B-0941}.  This was followed by application of the \textsc{photred} package as described in detail by \citet{nidever17}. \textsc{photred} efficiently runs the \textsc{daophot} photometry programs \citep{stetson87,stetson94} automatically on a night-by-night basis, after choosing options via a setup file. For the individual images, photometry was performed using the \textsc{allstar} program.  Astrometry is referenced to the system of the \textit{Gaia} catalog.  We also ran a separate processing where all four nights were grouped together and the forced photometry program \textsc{allframe} used to produce a single photometry list for each pointing. 

\begin{figure*}
\includegraphics[width=0.33\textwidth]{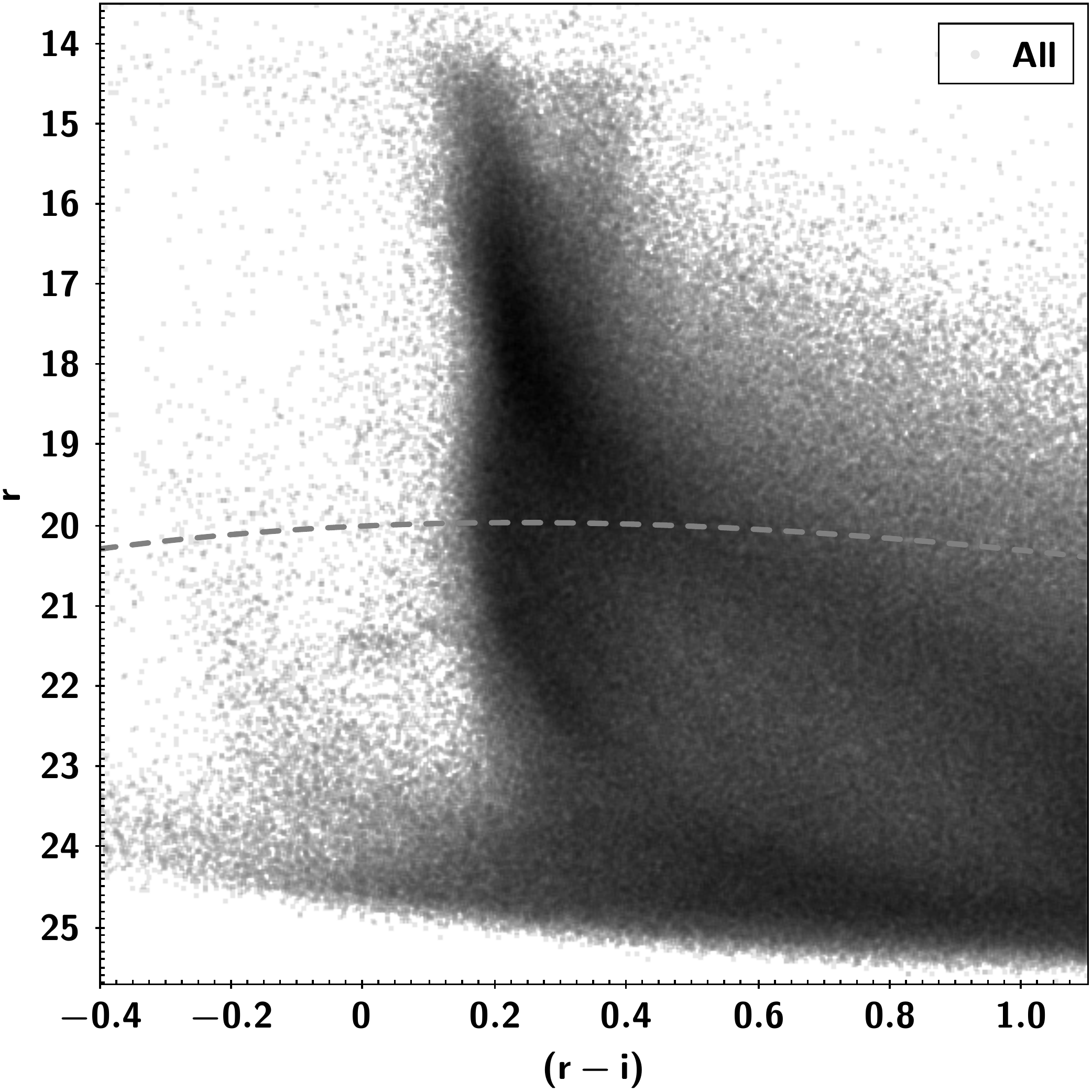}
\includegraphics[width=0.33\textwidth]{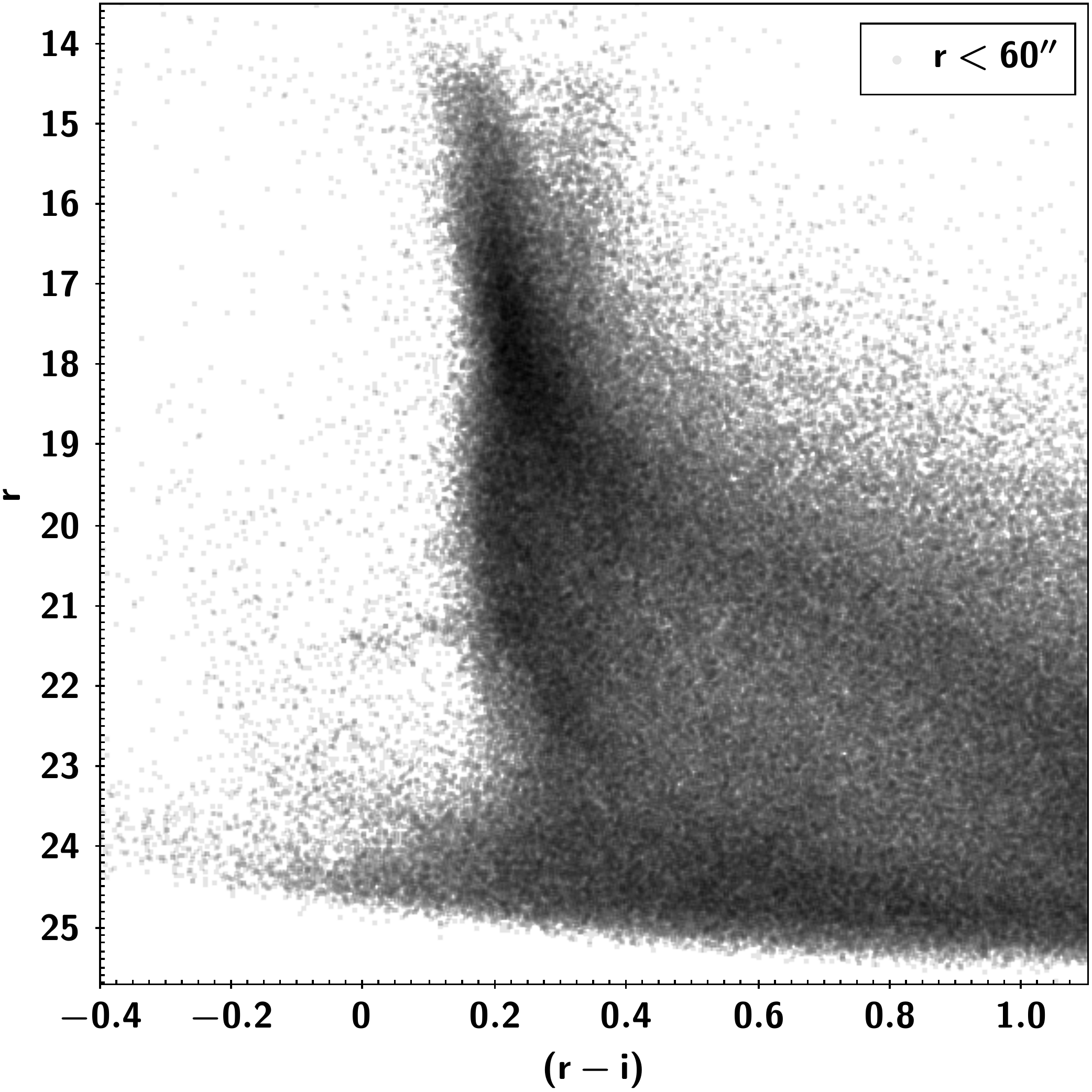}
\includegraphics[width=0.33\textwidth]{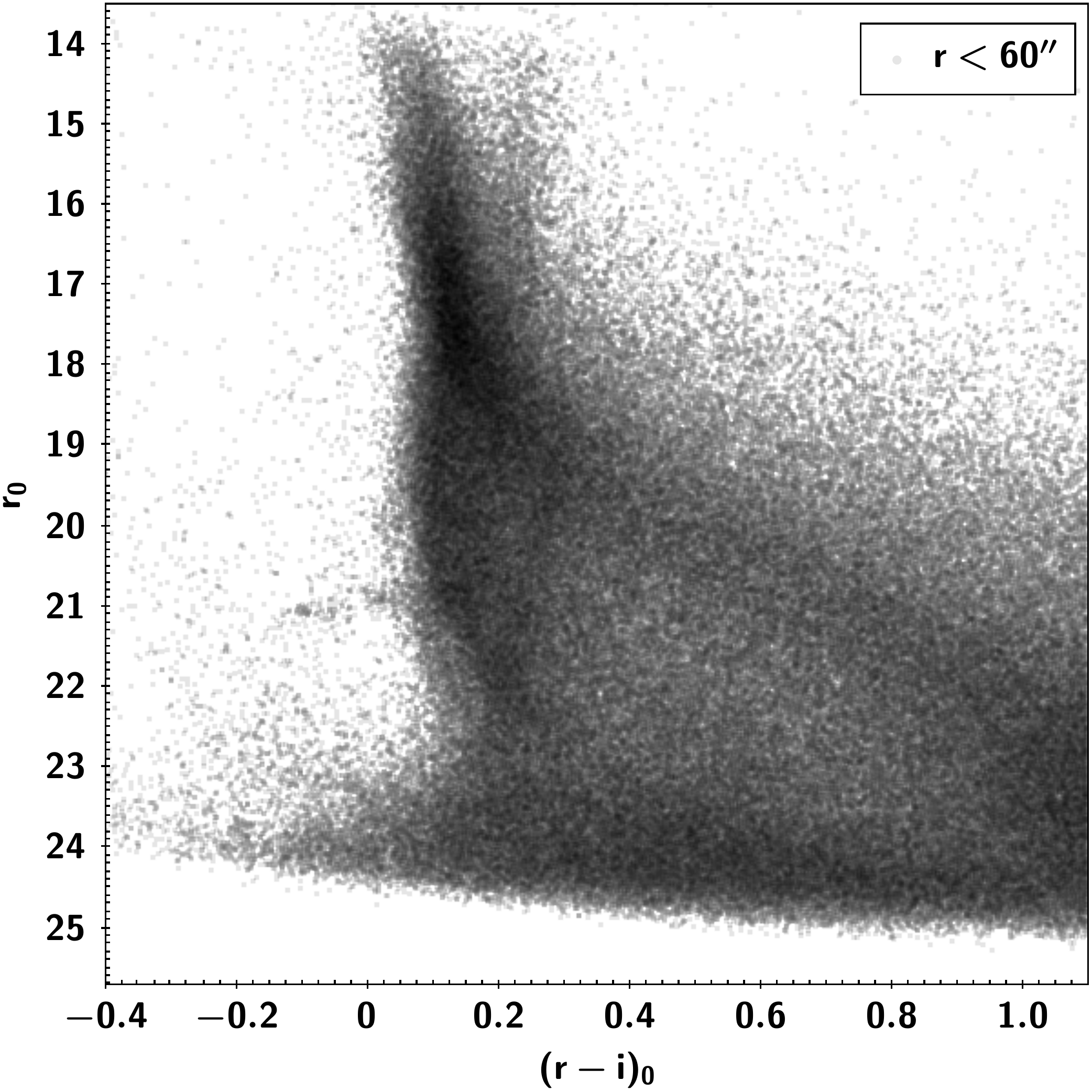}
\caption{(Left) Observational CMD of the full observed area in Ant 2, containing $\sim$ 1 million stars. The grey dashed line indicates the Gaia limit ($G=20$) we used to explore the blue plume population seen at $(r-i) \sim -0.2$. (Middle) Observational CMD showing only stars within $1\degr$ from the center of Ant 2, 297064 stars. (Right) Reddening-corrected CMD of stars within $1\degr$ from the center of Ant 2.}
\label{fig:CMD}
\end{figure*}

Photometric calibration was obtained from the standard star fields observed during clear conditions. The resulting photometry is on the SDSS system. The left color-magnitude diagram (CMD) shown in Figure~\ref{fig:CMD} contains 1,036,351 objects after isolating point sources by requiring CHI$\leq 3.0$, $-1.0\leq \rm{SHARP} \leq 1.0$, and PROB$\geq 0.8$. There are no clear features of the Ant 2 system in this CMD except by a Horizontal Branch (HB) at $r\sim 21$. The CMD is dominated by foreground populations, which at these low galactic latitudes ($b\sim 11\degr$) is large. The HB looks extended in brightness which is a likely consequence of both differential extinction and distance gradient within the galaxy, as we discuss later in this paper (see \S~\ref{sec:distance}). There is a hint of a blue plume population at $(r-i) < 0$ and $18 < r < 23.5$. Just from the CMD it is not obvious if this blue plume is evidence of a younger population present in Ant 2, or if it is formed by Galactic foreground stars. To explore this we analyzed parallaxes by Gaia EDR3 \citep{gaiaedr3} for stars brighter than $G=20$, shown in Figure~\ref{fig:CMD} (left panel) as stars above the grey dashed line. The blue population only shows up for a high parallax selection, and the resulting absolute magnitudes suggest these blue stars are mostly white dwarfs and probably some more exotic objects like extreme horizontal branch stars or extremely low mass white dwarfs, which are significantly brighter than normal white dwarfs. Although this quick analysis is limited only to the bright part of the CMD, the coherence of the observed feature suggest all of them are MW white dwarfs, which is not surprising since the line of sight to Ant 2 goes trough the Galactic disk. This conclusion does not preclude Ant 2 of having some young main sequence or Blue Straggler stars, but if these indeed exist they are likely few and hard to pick from the CMD.

The HB looks more defined and narrow when looking only at stars within a radius $r<1\degr$ (Figure~\ref{fig:CMD}, middle panel), where the extinction is more uniform and the distance gradient is less noticeable. Only when the CMD is corrected by extinction, a clear RGB and HB is visible (Figure~\ref{fig:CMD}, right panel). We obtained E(B-V) from the \citet{schlegel98} interstellar dust maps using the Python task \verb|dustmaps| \citep{green18}. The photometry is not deep enough to reach to the main sequence turnoff of Ant 2.

\subsubsection{Variable Stars} \label{sec:variable}

\begin{figure}
\plotone{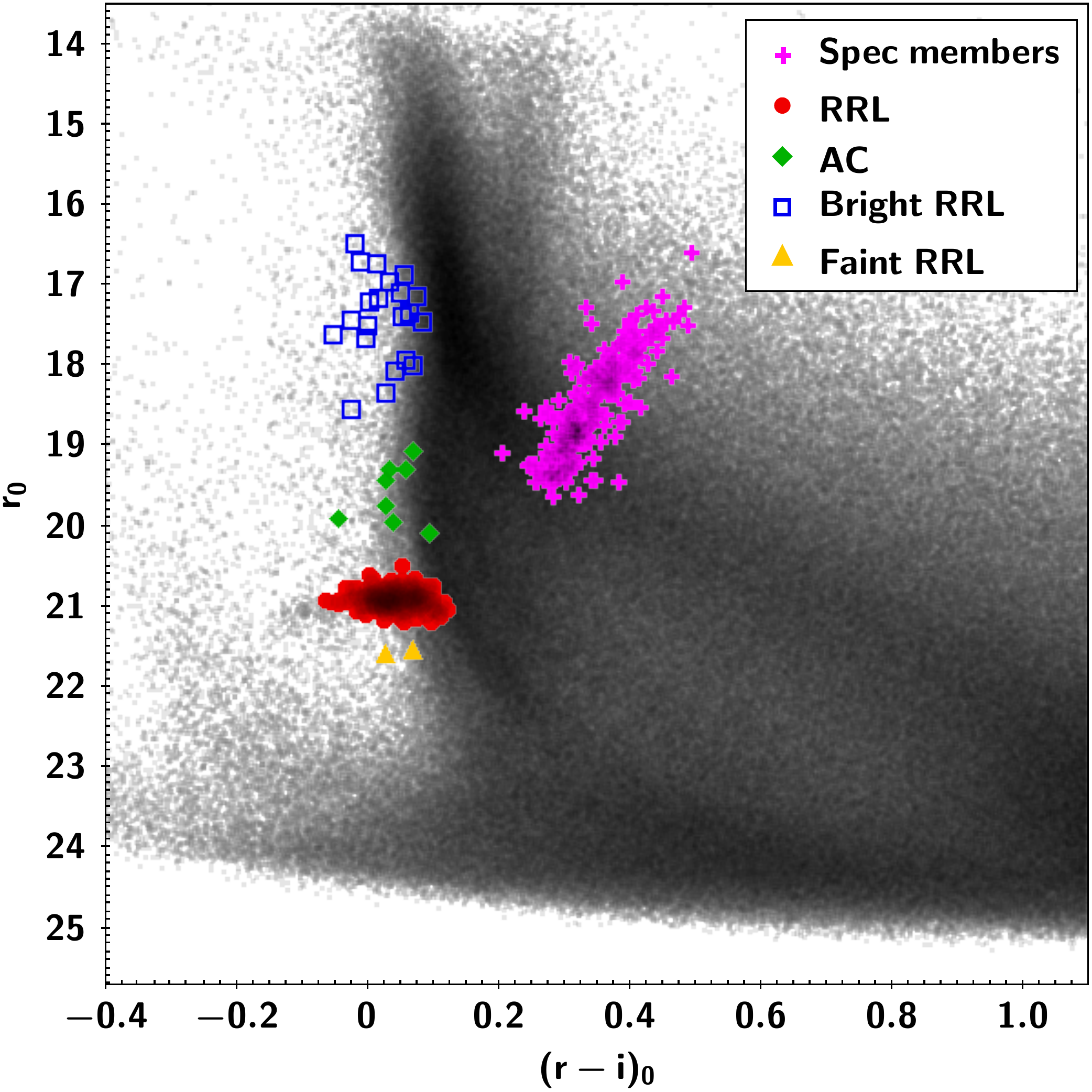}
\caption{Reddening-corrected CMD with the variable stars identified in this work. For reference we also show the high probability spectroscopic members in the RGB of Ant 2, as identified by \citet{Ji21}.}
\label{fig:CMDvar}
\end{figure}

\begin{figure}
\plotone{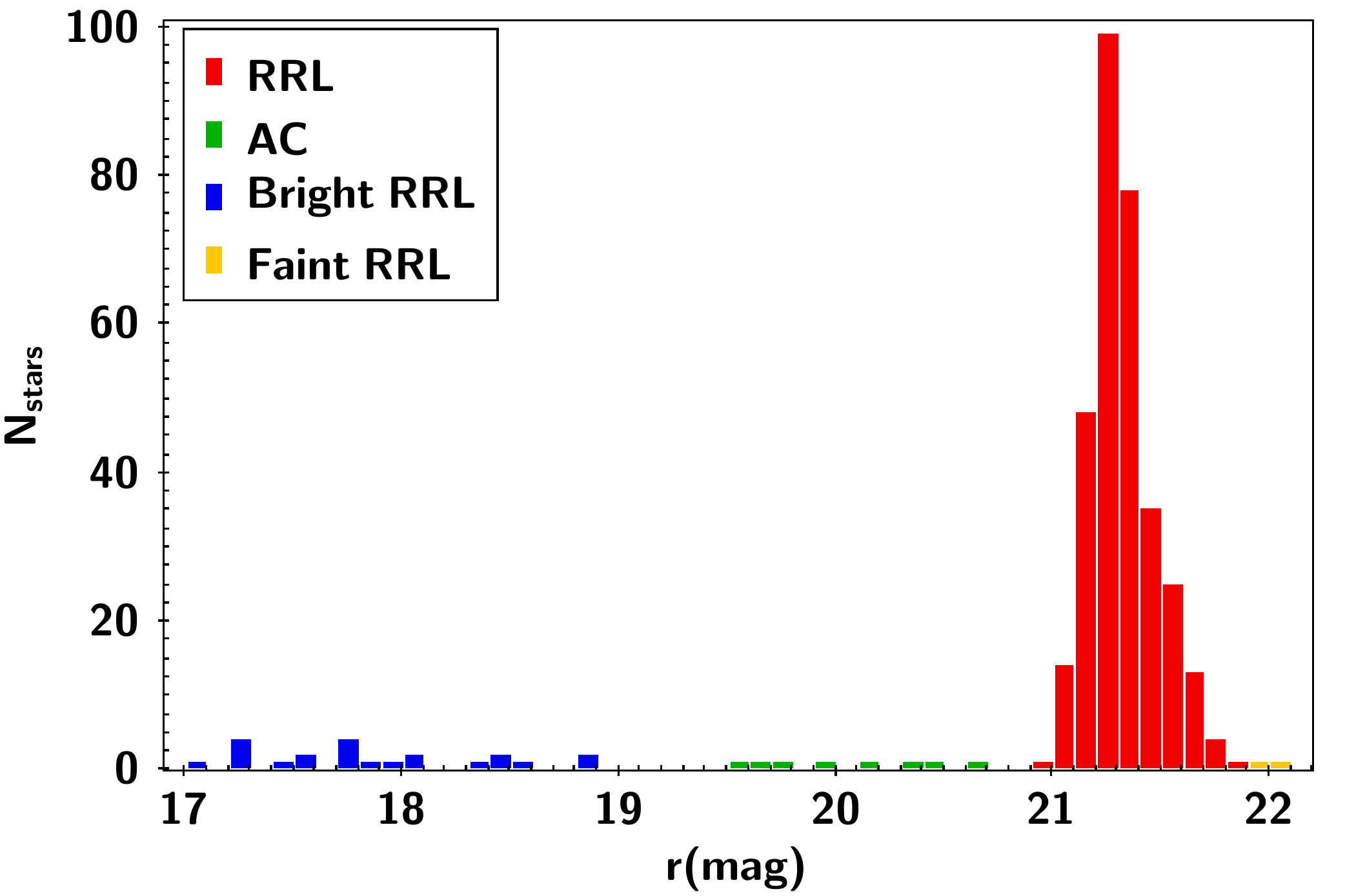}
\caption{Histogram of the mean $r$ magnitudes of all periodic variable stars found in the field of Ant 2.}
\label{fig:histogram_r}
\end{figure}

\begin{figure*}
\includegraphics[width=0.92\textwidth]{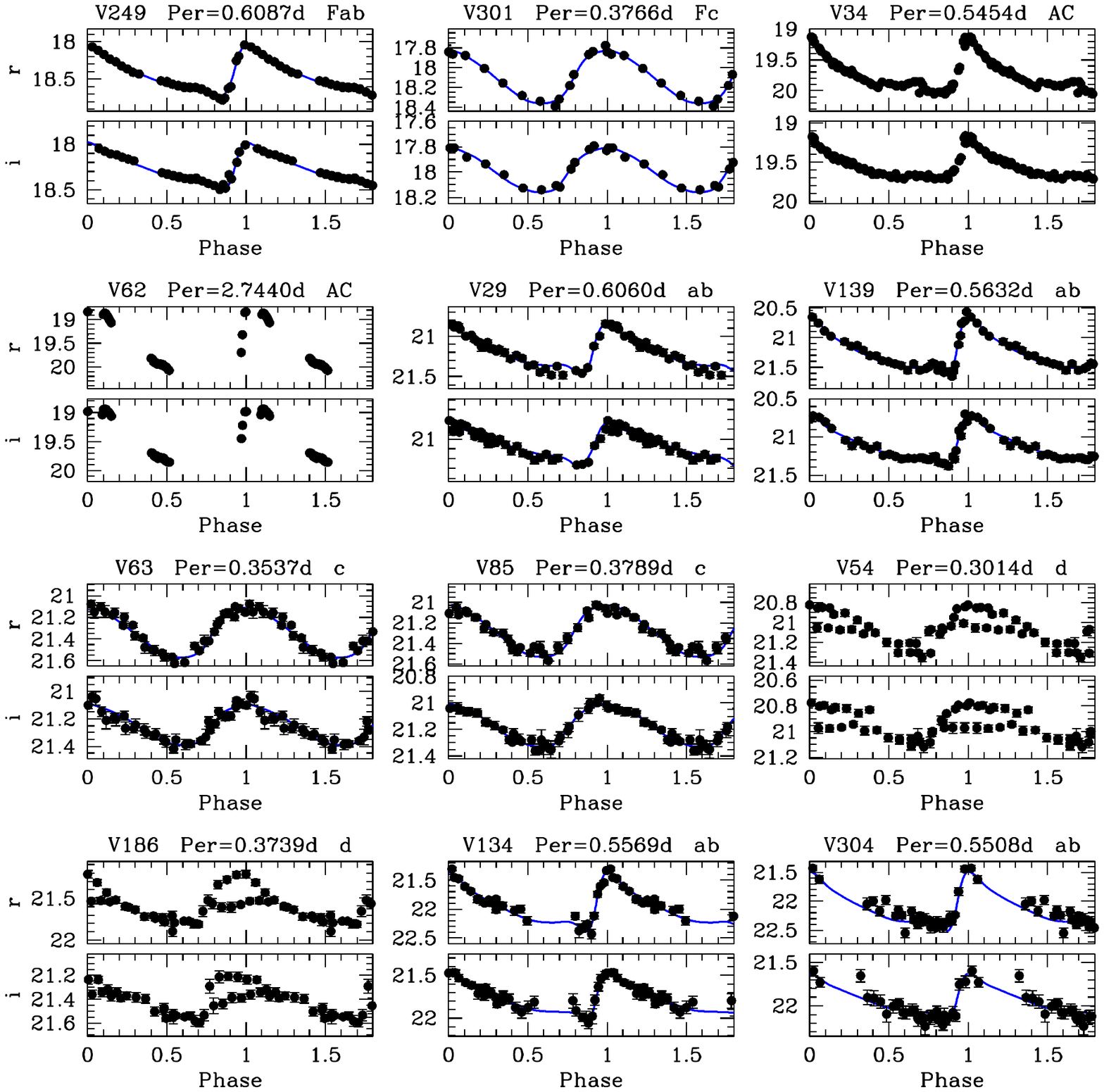}
\vspace{-4.8cm}
\caption{Sample of light curves. For each star the $r$, $i$ data are displayed in the top/bottom panels. Stars V249 and V301 are bright (foreground) RRL stars; V34 and V62 are AC; V29 and V139 are \rrab{} in Ant 2; V63 and V85 are \rrc{} in Ant 2; V54 and V186 are \rrd{} in Ant 2, phased with their first overtone periods; V134 and V304 are the two faint RRL stars in the background of Ant 2. For all the types $ab$ and $c$ the underlying blue line is the best fitted template to the light curve.}
\label{fig:lc}
\end{figure*}

We selected variable stars from our catalog following the same recipe as in previous works by our group, namely \citet{vivas16,vivas19,vivas20}. In particular for this work, we pre-selected stars which were flagged as variables in both the $r$ and $i$ band, have at least 12 observations in each band, have magnitudes in the range $17<r<22.5$, colors in the range $-0.3< (r-i) < 0.4$, and an amplitude in the $r$ band of at least 0.1 mag. Then, the pre-selected stars were searched for periodicity in the range $0.1-2.5$ days, with the final selection and classification was made by visual inspection of the phased light curves. These constraints were designed to find RRL and anomalous Cepheids (AC) in the region of the instability strip, with a very loose color constraint to give room to variables with high reddening. The light curves we examined have on average 33 epochs in each filter; in the region of overlap between fields there are light curves with as many as 68 epochs/filter.

We discovered numerous eclipsing binaries, a consequence of the large disk foreground population in our data, which are not studied further in this paper. Stars identified as first-overtone RRL pulsators, or \rrc{} stars, which are prone to be contaminated by eclipsing binaries of the W Ursa Major type \citep[e.g.][]{kinman10} because of their near to sinusoidal light curves, were carefully examined by looking at their colors and their location in a period-amplitude diagram. Several candidates were discarded as {\rrc } because either they were too red for being this type of pulsator or because their amplitudes and periods did not lie in the correct locus of a Bailey (period versus amplitude) diagram.

In total we found 350 periodic variable stars in the Ant 2 field which were preliminarily classified as {\rrab } (211 stars), {\rrc } (115 stars), {\rrd } (21 stars), and AC (3 stars). In the latter group the periods were too long ($>1$ day) for being classified as RRL. By placing the variable stars in the CMD (Figure~\ref{fig:CMDvar}), it seems clear that all the variables in the range of $19.5<r<20.8$ should be AC stars in Ant 2.  In this group there are eight stars, including the three variables with longer periods mentioned before as well as five stars classified initially as {\rrab } and {\rrc }. So after reclassifying these as AC stars, the final numbers are 341 RRL stars (207 \rrab, 114 \rrc, 21 \rrc), and eight AC. Table~\ref{tab:var} contains the properties of all the stars: ID, right ascension, declination, pulsational period, number of observations, amplitude and mean magnitude in each band ($r$ and $i$), extinction in $i$, and type of variable. The ID numbering from V1 to V350 was assigned according to the elliptical distance (see~\ref{sec:profile}), with the lower numbers being those closer to the center of Ant 2. Amplitudes and mean magnitudes for \rrab{} and \rrc{} were obtained from the best fitted template \citep[from the library by][]{sesar10} to the light curves. The mean magnitudes were obtained by integrating the templates in intensity units and transforming the result back to magnitudes.

As shown in Figure~\ref{fig:CMDvar} and in a histogram of magnitudes of the variable stars (Figure~\ref{fig:histogram_r}), there is a group of 22 bright RRL stars with $r<19.0$ which are classified as a foreground population from the halo, and possibly alternatively from the thick disk, of the MW. These bright RRL stars are located at distances between 16 kpc and 40 kpc from the Sun. The majority of the RRL stars in our sample (318 stars, 91\%), however, are concentrated in the HB of Ant 2 at $r\sim 21.3$ in the CMD. We discuss these stars in detail in the next section. In addition, there are two RRL stars (V134 and V304) which lie well below the HB ($r>21.9$), and thus appear to be located behind Ant 2 (see \S~\ref{sec:distantRR} for further discussion).

A sample of light curves is shown in Figure~\ref{fig:lc}, including two bright (foreground) RRL stars, two AC, six RRL stars in Ant 2 (two of each type), and finally the two faint RRL stars likely located behind Ant 2. The full photometry dataset and light curves for all the variables can be found in the appendix (\S~\ref{sec:data}). For the RRL stars in Ant 2 the typical photometric errors of the individual epochs are $0.04$ and $0.03$ mag in $r$ and $i$ respectively

Most of the periodic variable stars found in this work are new discoveries since they are fainter than the limiting magnitude for most wide surveys of variables such as the CRTS \citep{drake14} or Gaia \citep{clementini19}. We matched our catalog with the Gaia DR2 RRL star listing \citep{clementini19} and found 17 stars in common, all of them brighter than $r<20.3$. The RRL stars in Ant 2 are just too faint for Gaia. On the other hand, there are 39 Gaia RRL stars within our footprint, and we did not recover 21 of them, mostly because they were too bright ($r<17$) for our survey. There are, however, two Gaia RRL stars in the range $18.5<r<20.2$ that are in our magnitude range; they were indeed flagged as variable but we were unable to find a period for them\footnote{Gaia IDs are 5437028340146753664 and 5436057883696115328}. In one case the star was located close to the border of one field and had only $\sim 15$ epochs/band, which was not enough to obtain a period, although our data is consistent with the \textit{Gaia} period. In the second case, the \textit{Gaia} star has $G=20.2$, which is close to the faint limit for the RRL stars in that catalog. Our data does not agree either with the period nor with the amplitude given by \textit{Gaia}, hinting for a possible misclassification. 

\begin{deluxetable*}{rccccccccccccr}
\tabletypesize{\footnotesize}
\tablecolumns{12}
\tablewidth{0pc}
\tablecaption{Periodic Variable stars in the field of Ant 2 \label{tab:var}}
\tablehead{
ID & $\alpha$ & $\delta$ & Period & N (r) & Amp (r) & Mean r & N (i) & Amp (i) & Mean i & A$_i$ & Dist & $\sigma$Dist & Type\tablenotemark{a}\\
  & (deg) & (deg) & (d) &     & (mag)    & (mag) &        & (mag) & (mag) & (mag) & (kpc) & (kpc) & \\
}
\startdata
  V1 & 143.87486 & -36.91461 & 0.66093 & 41 & 0.61 & 21.29 & 41 & 0.38 & 21.16 & 0.27 & 129.4 &  2.6 &  ab \\
  V2 & 143.73833 & -36.86352 & 0.35250 & 65 & 0.66 & 21.10 & 67 & 0.41 & 20.98 & 0.26 &   --- &  --- &   d \\
  V3 & 143.90192 & -36.93597 & 0.39555 & 46 & 0.52 & 21.33 & 46 & 0.25 & 21.18 & 0.27 & 124.2 &  2.5 &   c \\
  V4 & 143.73605 & -36.91526 & 0.56934 & 26 & 0.88 & 21.31 & 25 & 0.66 & 21.19 & 0.26 & 127.2 &  2.5 &  ab \\
  V5 & 143.93247 & -36.79384 & 0.39872 & 67 & 0.49 & 21.37 & 66 & 0.30 & 21.25 & 0.26 & 129.2 &  2.6 &   c \\
  V6 & 143.94381 & -36.98518 & 0.60702 & 33 & 0.28 & 21.36 & 34 & 0.21 & 21.19 & 0.28 & 128.4 &  2.6 &  ab \\
  V7 & 143.83054 & -36.99803 & 0.58207 & 34 & 0.63 & 21.34 & 33 & 0.52 & 21.19 & 0.28 & 127.1 &  2.5 &  ab \\
  V8 & 143.83305 & -37.02513 & 0.37702 & 34 & 0.49 & 21.40 & 29 & 0.30 & 21.24 & 0.28 & 126.2 &  2.5 &   c \\
\enddata
\tablenotetext{a}{Types: ab $=$ type ab RRL stars; c $=$ type c RRL stars; AC $=$ anomalous Cepheids; Fab $=$ type ab field RRL stars; Fc $=$ type c field RRL stars;}
\tablecomments{Table~\ref{tab:var} is published in its entirety in the machine-readable format. A portion is shown here for guidance regarding its form and content.}
\end{deluxetable*}

\section{RR Lyrae stars in Ant 2} \label{sec:rrl}

There are 318 RRL stars in Ant 2 of which 193 are \rrab, 104 are \rrc, and 21 are double mode pulsators, or \rrd. We discuss their properties in the following subsections:

\subsection{Oosterhoff Type} \label{sec:oo}

\begin{figure}
\plotone{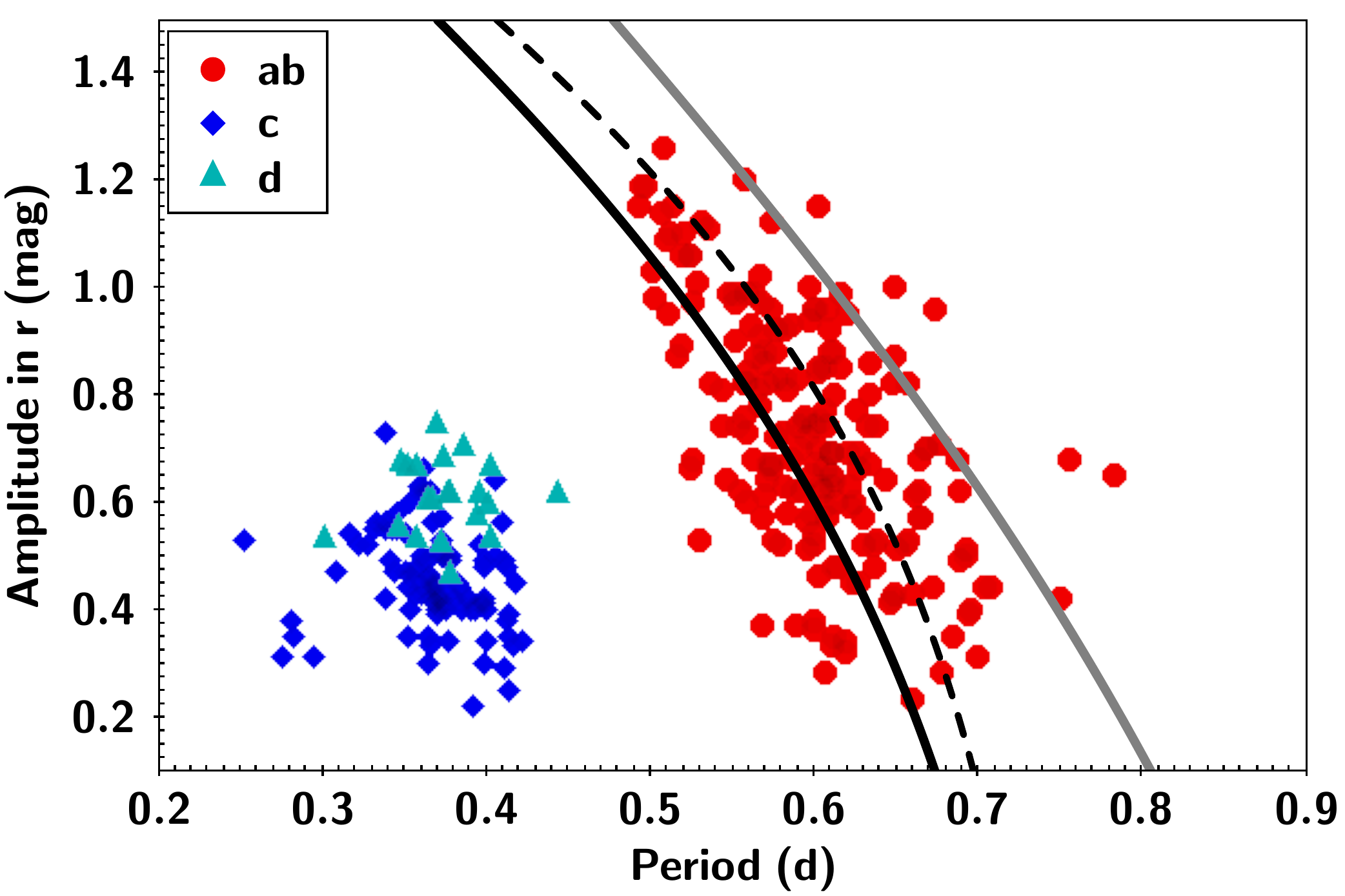}
\caption{Bailey diagram of RRL stars in Ant 2. The black solid, grey solid and black dashed lines represent the locus of Oo I, Oo II and Oo-int systems, scaled to the r-band from \citet{fabrizio19}.}
\label{fig:Bailey}
\end{figure}

The Oosterhoff type \citep[Oo,][]{oosterhoff39} was originally applied to RRL in galactic globular clusters; it revealed a dichotomy in the mean periods and in the ratio of the numbers of fundamental (\rrab) and first-overtone (\rrc) variables. This dichotomy is not observed in most of the satellite galaxies of the MW \citep{catelan09} and M31 \citep{martinez17}, which are usually classified as Oosterhoff-intermediate (Oo-int) since their properties lie in between those for the two original Oo I and Oo II groups. 

The mean period of the {\rrab } in Ant 2 is 0.599 days, while the {\rrc } have a mean period of 0.368 days. The ratio $N_c / (N_{ab}+N_c)$ is 0.35. From \citet{smith95}, mean periods of {\rrab } are 0.55d and 0.64d, and mean periods of {\rrc } are 0.32d and 0.37d for Oo I and Oo II systems, respectively. From the same source, the ratio of type $c$ stars over the sum of $ab$ and $c$ are 0.17 and 0.44 for Oo I and Oo II, respectively. Comparing these values with our data, Ant 2 is clearly in the middle between the two Oo groups, suggesting a Oo-int classification the same as found for many other dSph galaxies in the MW neighborhood.  

A Bailey diagram, a plot of amplitude of the light curve versus period, is now considered \citep{smith95} the best way to infer the Oo classification of a stellar system. In Figure~\ref{fig:Bailey} we show the Bailey diagram for Ant 2, and the fiducial lines for Oo I and Oo II as defined by \citet{fabrizio19} but scaled by a factor of 0.94 to take into account that we have amplitudes in the $r$ band and not in $V$. The scale factor was calculated by comparing the amplitudes of RRL stars in common between SDSS Stripe 82 \citep{sesar10} and CRTS \citep{drake14}. The {\rrab } stars in Ant 2 are mostly found between the fiducial lines for the Oo I and Oo II group, confirming that Oo-int is indeed the best classification for this system. 

We estimate the percentage of Oo I and Oo II \textit{like} stars in Ant 2 by counting how many \rrab{} stars are located on each side of the Oo-int relation by \citet{fabrizio19}. We obtained that 75\% of the stars are Oo I while 25\% are Oo II \textit{like}. These results are consistent with what was found in \citet{martinez17} for RRL stars in Local Group dwarf galaxies where the majority of stars ($\sim$ 80\%) are distributed close to the locus of Oo I type.

\subsection{Spatial Distribution} \label{sec:distribution}

\begin{deluxetable*}{ccccccc}
\tablewidth{0pc}
\tablecaption{Shape parameters of Ant 2\label{tab:ellipse}}
\tablehead{
  & $\alpha$(J2000.0) & $\delta$ (J2000.0) & $\epsilon$ & PA & $1\sigma$ semi-major axis & $r_h$ \\
    & (deg) & (deg) & & (deg) & (deg) & (deg) \\
}
\startdata
This work &  $143.83\pm 0.02$ & $-36.86 \pm 0.02$ & $0.28 \pm 0.03$ & $141\pm 3$ & $0.88 \pm 0.03$ &  \\
\citet{T19} & $143.87 \pm 0.05$ & $-36.77 \pm 0.10$ & $0.38 \pm 0.08$ & $156 \pm 6$ & & $1.27 \pm 0.12$ \\
\citet{Ji21} & $143.81 \pm 0.05$ & $-36.70 \pm 0.08$ & $0.60 \pm 0.04$ & $154 \pm 2$ & & $1.11 \pm 0.08$ \\
\enddata
\end{deluxetable*} 

\begin{figure}
\plotone{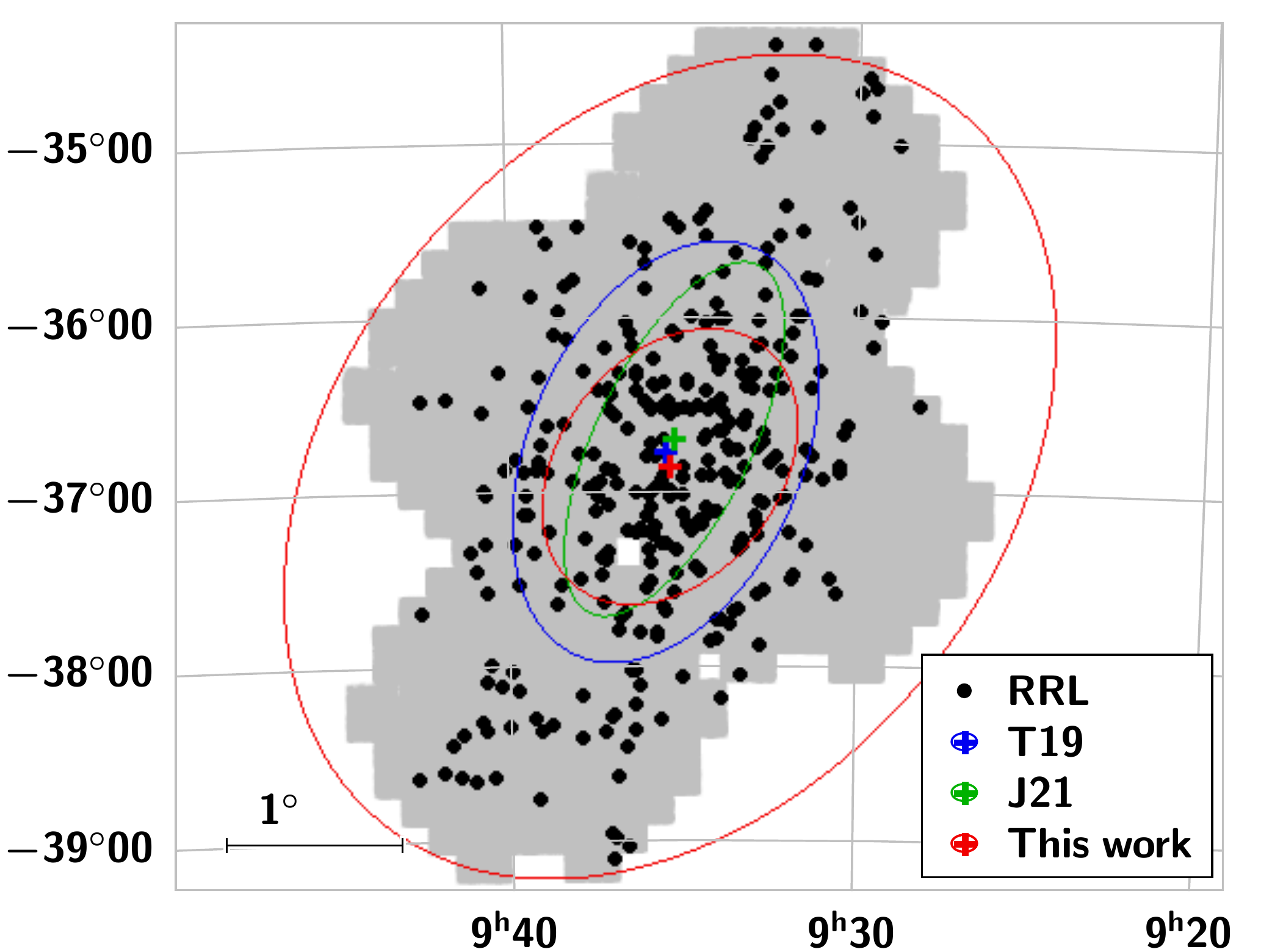}
\caption{Shape of the Ant 2 galaxy from fitting a bi-variate Gaussian to the RRL star sample. In red we show the center and the $1\sigma$ and $3\sigma$ semi-major axis of the resulting ellipse. For comparison we also show the center (blue and green crosses) and shape parameters (blue and green ellipses) determined by \citet[][T19]{T19} and \citet[][J21]{Ji21}. For those two cases the semi-major axis of the ellipse represents the half-light radius ($r_h$) of the galaxy.}
\label{fig:ellipse}
\end{figure}

The on-sky spatial distribution of the RRL stars in Ant 2 can be seen in Figure~\ref{fig:Sky}. RRL stars were found in the full footprint of our observations, which reaches $\sim 2r_h$. Although the density of stars clearly decreases with distance from the center of the galaxy, it is very likely that there are additional stars extending beyond $2r_h$.

Because of the very low surface brightness of Ant 2, combined with its low galactic latitude, contamination by fore/background stars is a challenge when analyzing the properties of the galaxy. By contrast, the RRL stars are a very pure sample of members of the galaxy. There is virtually no contamination by halo stars at this large distance from the Galactic center. Thus, we used the RRL stars to infer the shape of Ant 2 by fitting a bi-variate Gaussian distribution\footnote{using AstroML fit-bivariate\_normal routine \citep{ivezic14}} to the RRL stars to find the center, ellipticity ($\epsilon$), position angle (PA) and size of Ant 2. The results can be seen in Figure~\ref{fig:ellipse}, and in Table~\ref{tab:ellipse}. Within the $1\sigma$ errors, we found the same center and ellipticity as \citet{T19}. The position angle is slightly different with respect to the one obtained \citet{T19} but within $2\sigma$ errors of each other. There is more disagreement with the structural parameters found by \citet{Ji21}, particularly in the ellipticity, since they find a significantly more elongated structure than us. We obtain a $1\sigma$ semi-major axis of $52\farcm 6$. Since we fitted Gaussian distributions this value is not directly comparable with $r_h$ from either \citeauthor{T19} nor \citeauthor{Ji21}. We note that this elongation is aligned with the reflex corrected proper motions, once the LMC is accounted for \citep[see][for more details]{Ji21}.

\subsection{Distance to Ant 2 and Distance Gradient} \label{sec:distance}

\begin{figure}
\includegraphics[width=0.46\textwidth]{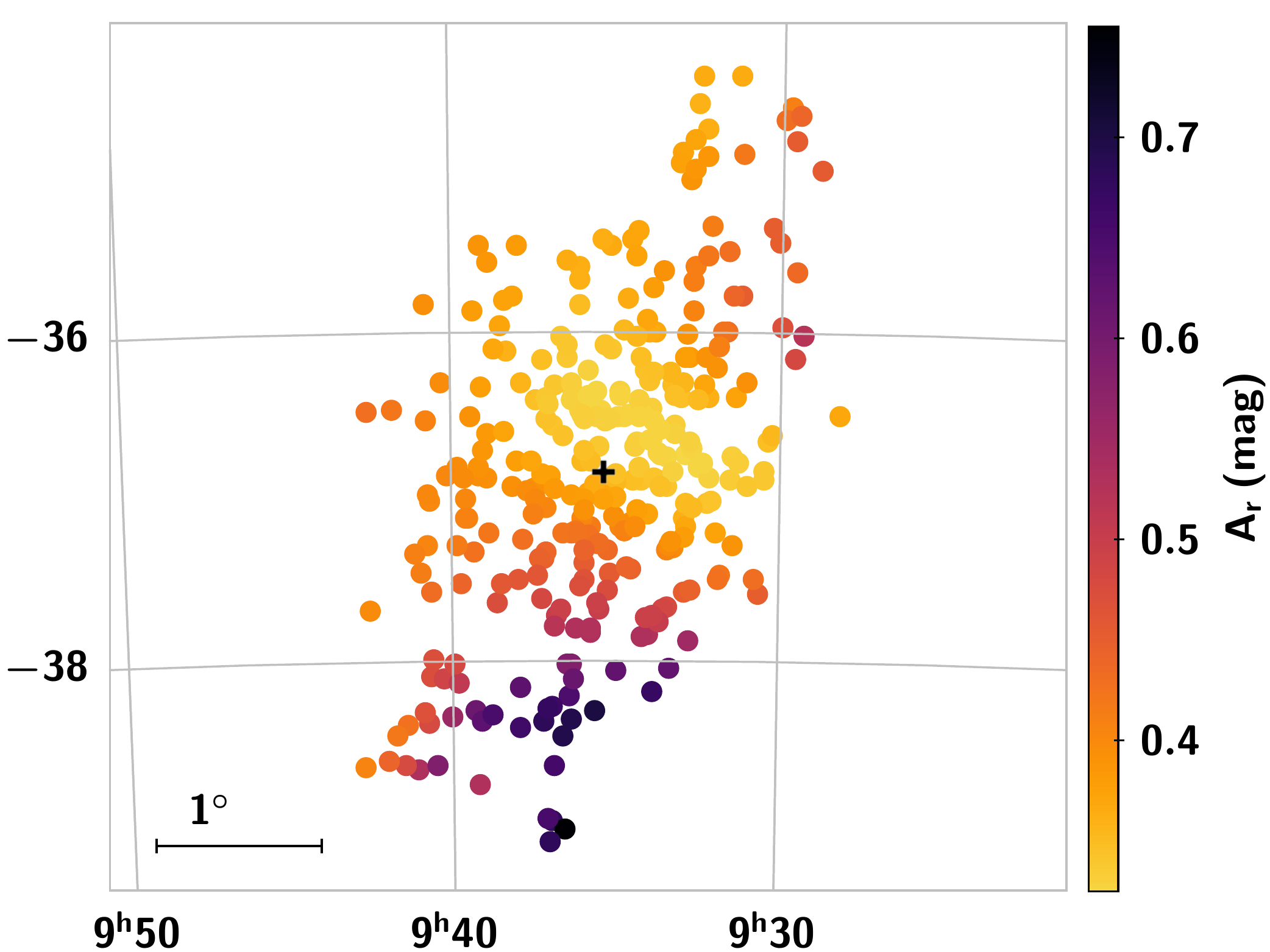}
\includegraphics[width=0.47\textwidth]{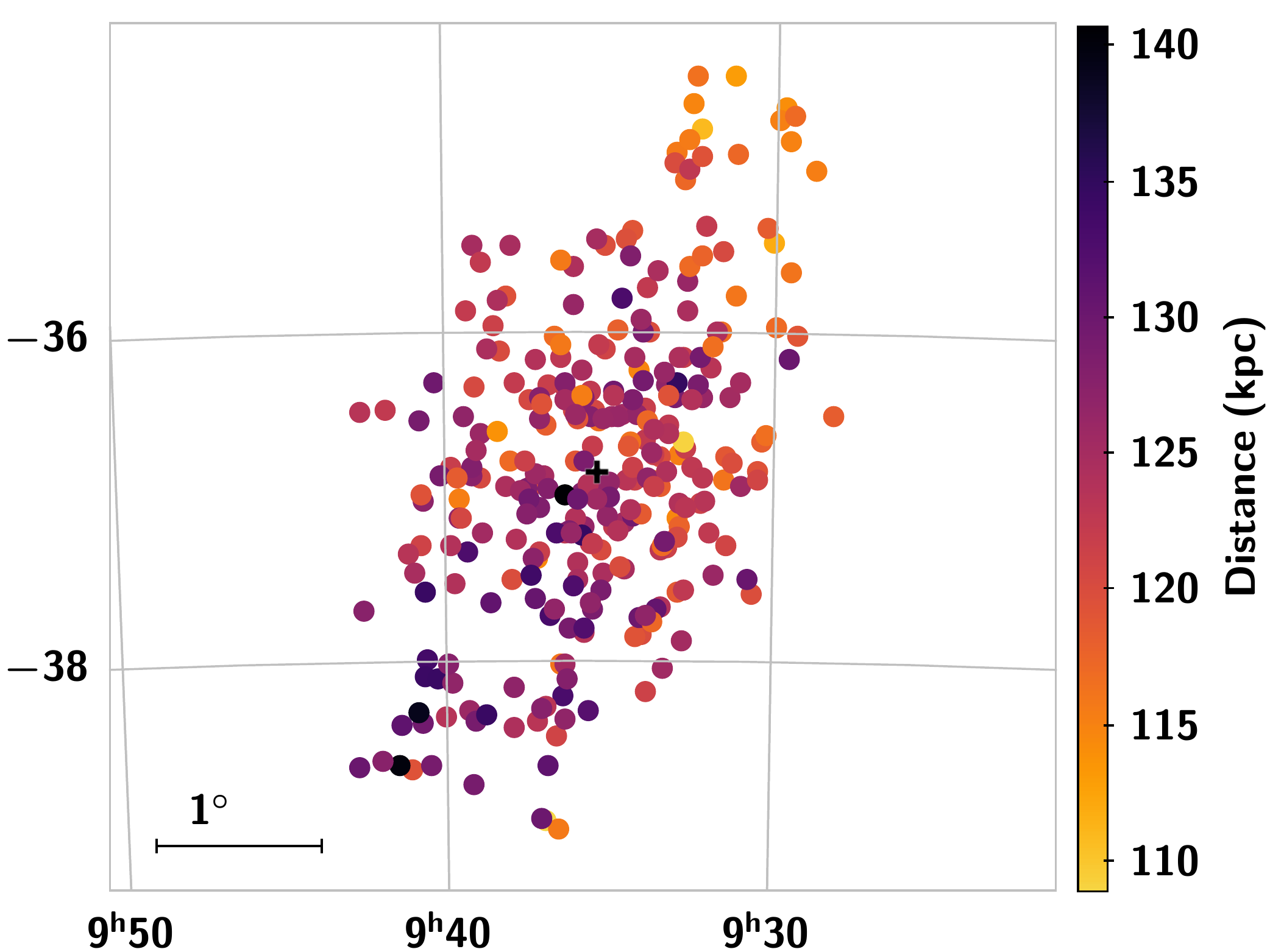}
\caption{(Top) Map in equatorial coordinates of the interstellar extinction in the $r$ band for all the RRL stars found in Ant 2 from \citet{schlafly11}. (Bottom) Map of the individual distances to RRL stars (type ab and c only). In both plots, the black cross marks the center of the distribution of the RRL stars. }
\label{fig:extinction}
\end{figure}

\begin{deluxetable}{lrrr}
\tablewidth{0pc}
\tablecaption{Distance to Ant 2\label{tab:distance}}
\tablehead{
Type & Distance (kpc) & Std Dev (kpc) & N \\
}
\startdata
ab & 124.7 & 5.5 & 193 \\
c & 123.0 & 4.9 & 104 \\
\hline
ALL & 124.1 & 5.4 & 297 \\
\enddata
\end{deluxetable}

\begin{figure*}
\includegraphics[width=0.49\textwidth]{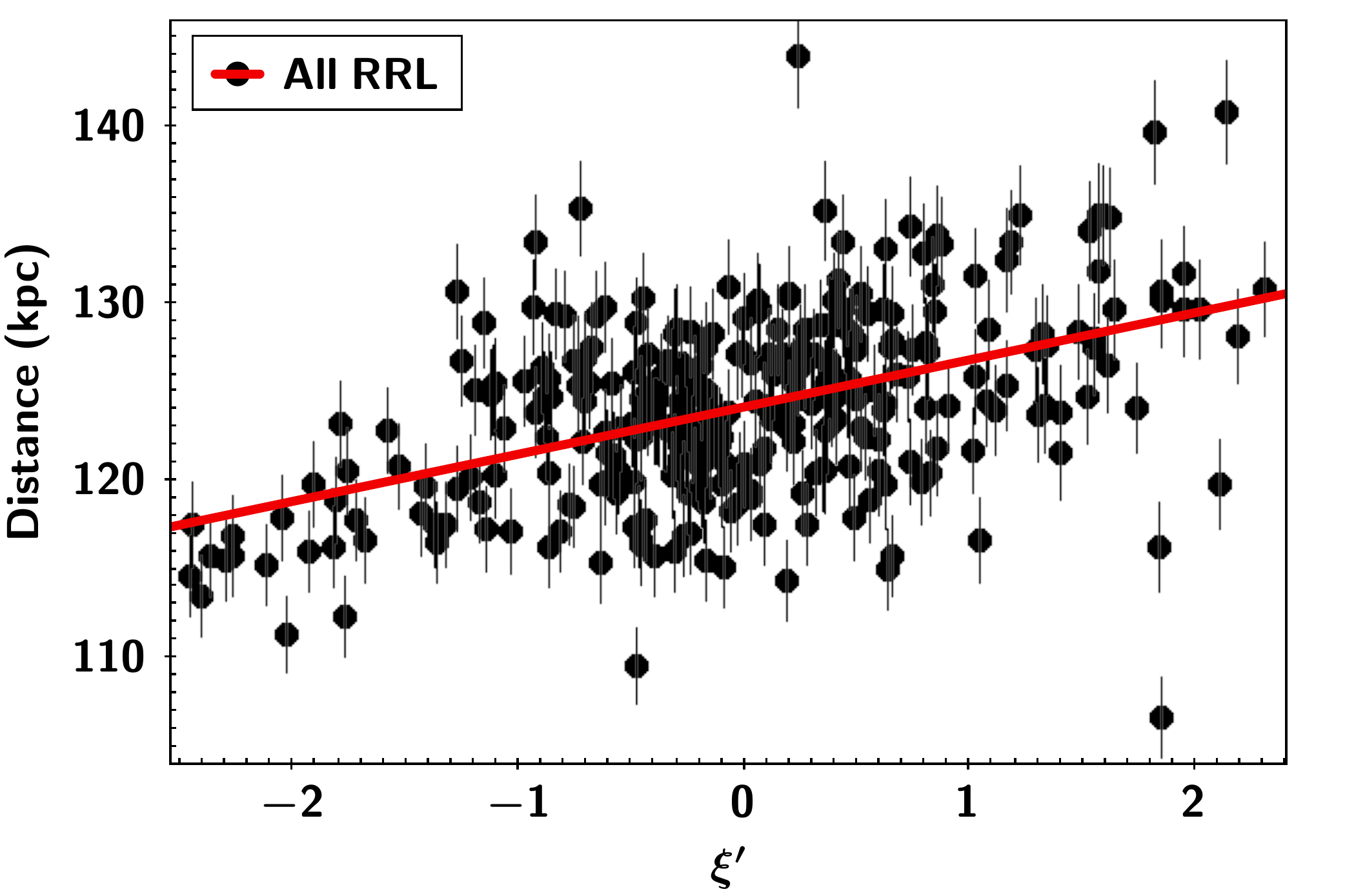}
\includegraphics[width=0.49\textwidth]{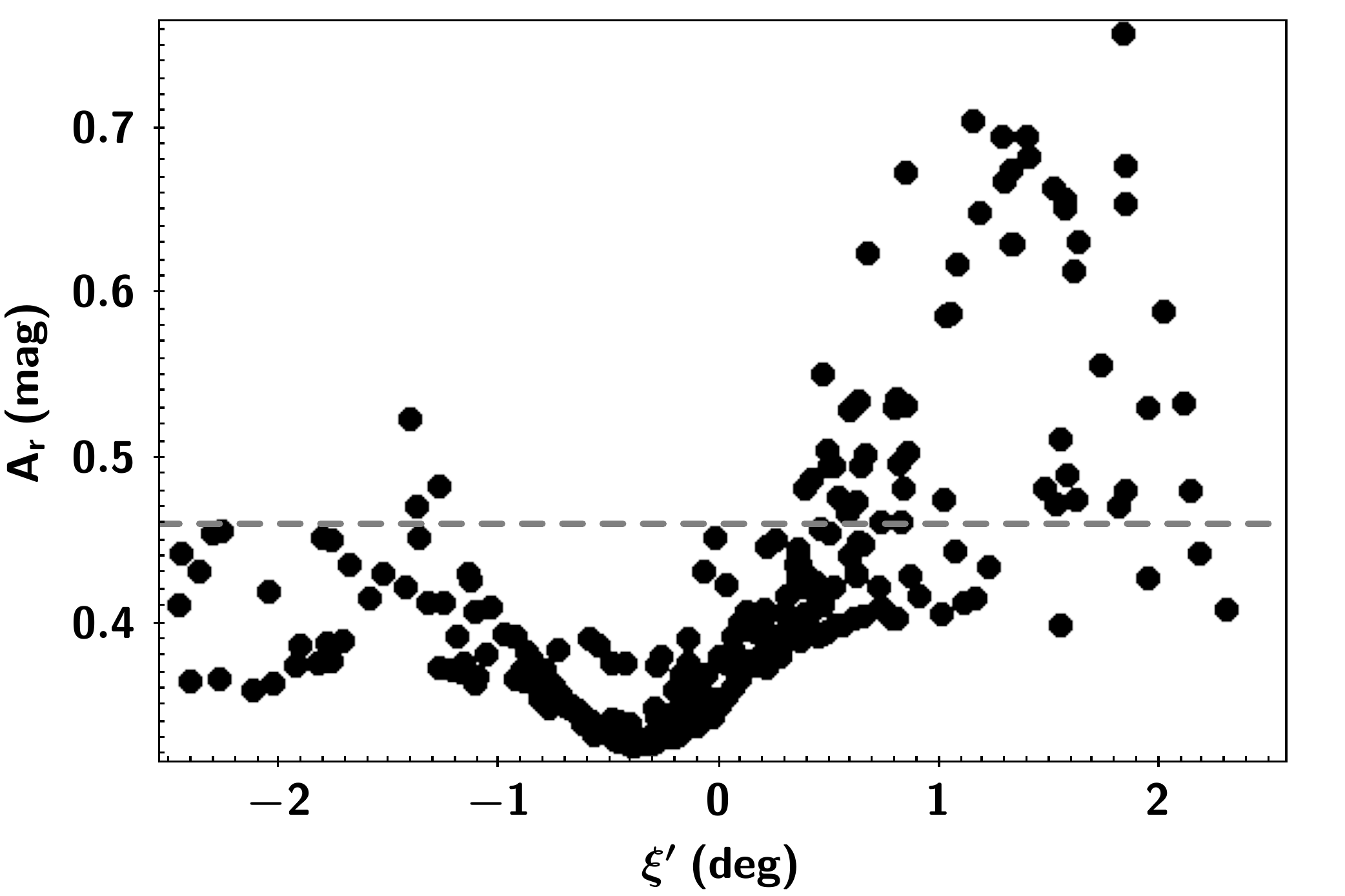}
\includegraphics[width=0.49\textwidth]{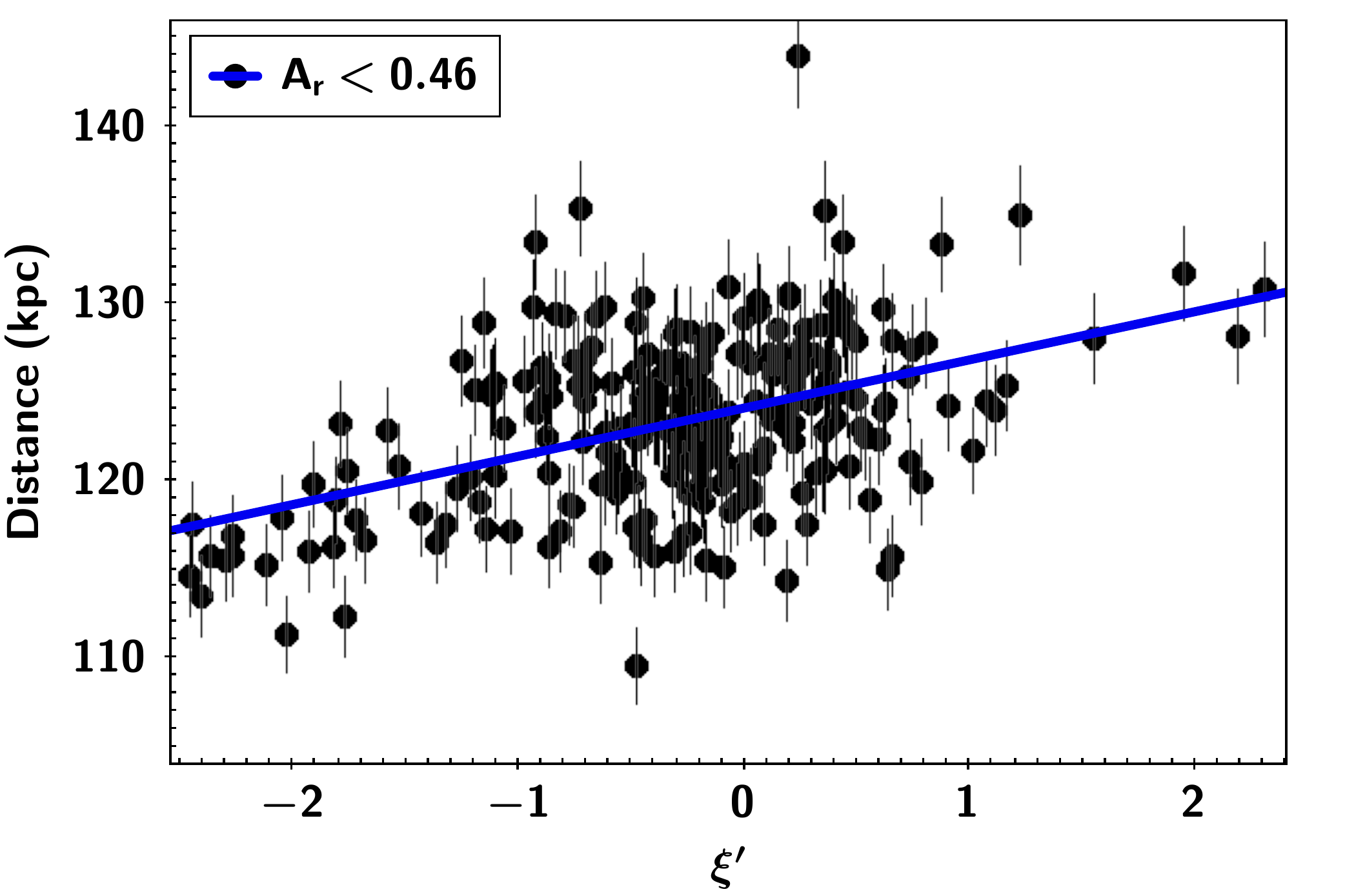}
\includegraphics[width=0.49\textwidth]{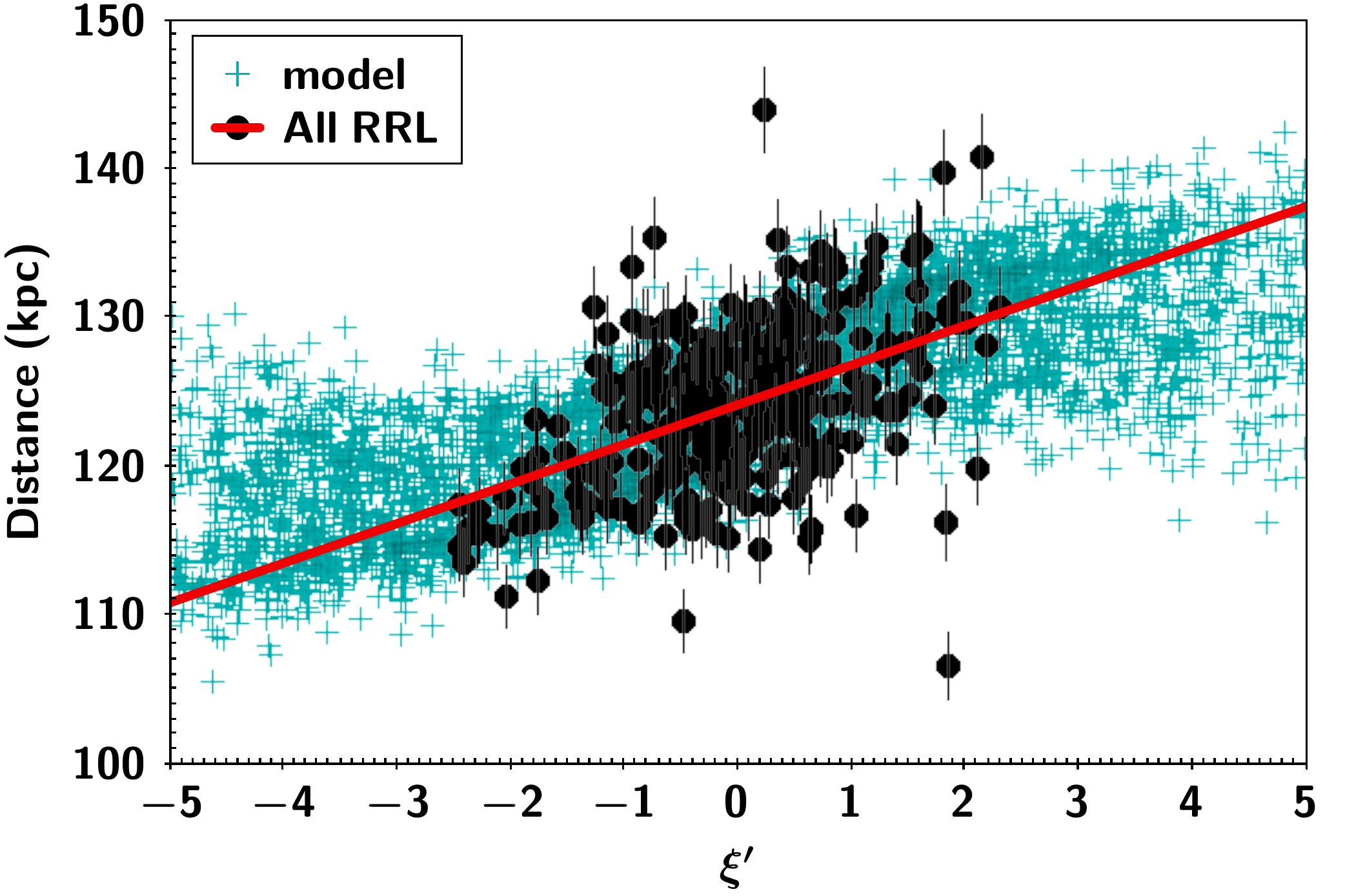}
\caption{(Left) Distance to the RRL stars along the semi-major axis of Ant 2. The red line is the best fit to the data. The bottom plot shows the same as above but for the RRL stars in the low extinction sample. (Right, top) Interstellar extinction, $A_r$, of RRL stars along the semi-major axis of Ant 2. The dashed grey line at $A_r=0.46$ is an arbitrary limit to define a low-extinction sample. (Right, bottom) Same as the left panels but with a wider view in both axis. The cyan crosses in the background represent the disruption model of Ant 2. The red line is the same as in the top left panel.}
\label{fig:distance_gradient}
\end{figure*}

One of the most important characteristics of RRL stars is that they are very good standard candles. Here we use the large number of variables in Ant 2 to determine the distance to the galaxy. Extinction is high and variable in this part of the sky due to the low galactic latitude location of Ant 2 (Figure~\ref{fig:extinction}). We obtained extinctions from the \citet{schlafly11}'s recalibration of the dust maps by \citet{schlegel98}. The mean extinction of the RRL stars in Ant 2 in the $r$ band is $A_r=0.41$ mag, but is as high as 0.76 mag for some stars. Figure~\ref{fig:extinction} shows that there is a region of relatively low extinction corresponding approximately with the central region of the galaxy. The extinction increases towards the extremities of Ant 2, particularly towards the South.

After correcting magnitudes for interstellar extinction, we calculated the distance to each individual RRL star using the Period-Luminosity (PL) relation by \citet{caceres08} in the $i$ band, which has a formal uncertainty of 0.045 mag. This relationship has a metallicity dependence for which we used the recent value of [Fe/H]$=-1.90$ (with a dispersion $\sigma=0.34$) derived by \citet{Ji21}. The PL relation was applied to all {\rrab } and {\rrc } stars (Table~\ref{tab:distance}). For the later, we first fundamentalized the periods. We did not use {\rrd } stars because for these stars we did not fit a template to the light curves and hence do not have a mean magnitude measured in the same way as the other stars. The results from  {\rrab } and {\rrc } are consistent with each other (Table~\ref{tab:distance}).

The mean distance given by the 297 RRL stars is 124.1 kpc, with a standard deviation of 5.4 kpc. The corresponding distance modulus is $\mu_0=20.47$ with a dispersion of 0.09 mag. The mean of the individual errors in Ant 2 is 2.5 kpc, and thus significantly lower than the standard deviation of the distribution. In order to ascertain what might be causing the large standard deviation in the distance distribution we first explored any spatial dependence of distance (Figure~\ref{fig:extinction}, bottom). It is clear from this map that there is a gradient of distance in Ant 2 with the South-East part of the galaxy being farther away than the North-West side. However, extinction is much higher to the South-East of Ant 2 (Figure~\ref{fig:extinction}, top) and may be influencing the distance determinations.

To better explore any possible correlation between the observed distance gradient and the extinction we transformed the equatorial coordinates ($\alpha$, $\delta$) to planar coordinates ($\xi$, $\eta$) with Ant 2 centered at ($0,0$), and we rotated the system so the semi-major axis of the galaxy obtained here (Table~\ref{tab:ellipse}) lies along the $x$-axis ($\xi'$). Figure~\ref{fig:distance_gradient} (top) shows the extinction as a function of the semi-major axis. In this coordinate system, the South-East part of Ant 2 is on the right side of the plot (positive $\xi'$). The middle panel shows the distance to the RRL stars as a function of the semi-major axis and the distance gradient suggested by the map in Figure~\ref{fig:extinction} (bottom) is now clearer. The red line shows the best linear fit to the data, which has a slope of 2.66 kpc/deg along the semi-major axis, and a distance at $\xi'=0\degr$ of 124.2 kpc. Thus, in 5 degrees from one side of the galaxy to the other, there is a difference in distance of $\sim 13$ kpc, which is much larger than the individual errors in distance. 

As seen in the upper panels of Figure~\ref{fig:distance_gradient}, the behaviour of both extinction and distance along the semi-major axis is completely different and thus the observed distance gradient is likely real rather than an effect of improper reddening correction. As a final check, we defined a ``low-extinction" sample defined (arbitrarily) as all stars with $A_r<0.46$ mag. The distance gradient is still visible in this reduced sample of 238 stars (Figure~\ref{fig:distance_gradient}, bottom left panel). The best fitted line in this case has a slope of 2.72 kpc/deg along the semi-major axis, and intersection with the y-axis at 124.1 kpc. These values are almost identical to the full sample and confirm that extinction is not causing the observed distance gradient. The gradient measured as a function of declination is $-2.44$ kpc/deg (full sample). We provide this alternative value since declination is a fixed parameter and the semi-major axis depends instead on the assumed structural parameters of the galaxy.

The bottom right panel in Figure~\ref{fig:distance_gradient} shows that the distance gradient observed in the RRL star population agrees quite well with the expectations from the disruption model of Ant 2 developed by \citet{Ji21} and modified here to account for the mean distance given by the RRL stars which is slightly different to the value assumed in \citet[][132 kpc]{Ji21}. More discussion on the disruption model can be found below in \S~\ref{sec:model}.

\subsubsection{RR Lyrae stars beyond Ant 2?}
\label{sec:distantRR}

We return to the case of stars V134 and V304, the two \rrab{} stars that lie below the HB of Ant 2, at mean $r$ magnitudes of 21.94 and 22.07 respectively. We showed their light curves in Figure~\ref{fig:lc}. They are well sampled on both filters and they have small photometric errors, casting no doubts on their classification. Following the same procedure to calculate distances discussed in the previous section we estimate that V134 is located at $167 \pm 4$ kpc and V304 is at $173\pm 4$ kpc. Within $1\sigma$ errors, both stars are located at the same distance, but they are $\sim 45$ kpc behind Ant 2, which is a distance much larger than the distance errors or the distance dispersion of Ant 2.

The stars are well within $2r_h$ of Ant 2, at $0\fdg 6$ and $1\fdg 4$ from the center of the galaxy, both of them to the South-West side of the field (Figure~\ref{fig:Sky}). They are separated by $47\arcmin$ from each other. 

By integrating the number density radial profile of RRL from \citet{medina18} between 150 and 200 kpc, we estimate only $0.2$ RRL stars are expected in that range of distances in an area of 12 sq. deg. (our surveyed area). Thus, the possibility of having two random halo field stars this close in the sky, and at this distance is very low. These stars are either an indication of a separate stellar system behind Ant 2, or they may be part of material that have been torn apart from Ant 2 in the past. Since the stars are separated by $47\arcmin$ from each other, it is unlikely they belong to a compact stellar system such as a UFD galaxy. However, in principle they could be members of a stream. Models of the disruption of Ant 2 (see \S~\ref{sec:model}) do not support the presence of debris material behind the main body of the galaxy. At present then, we do not have a good explanation for the origin of these two very distant RRL stars.

\subsection{Radial Profiles} \label{sec:profile}

\begin{figure}
\plotone{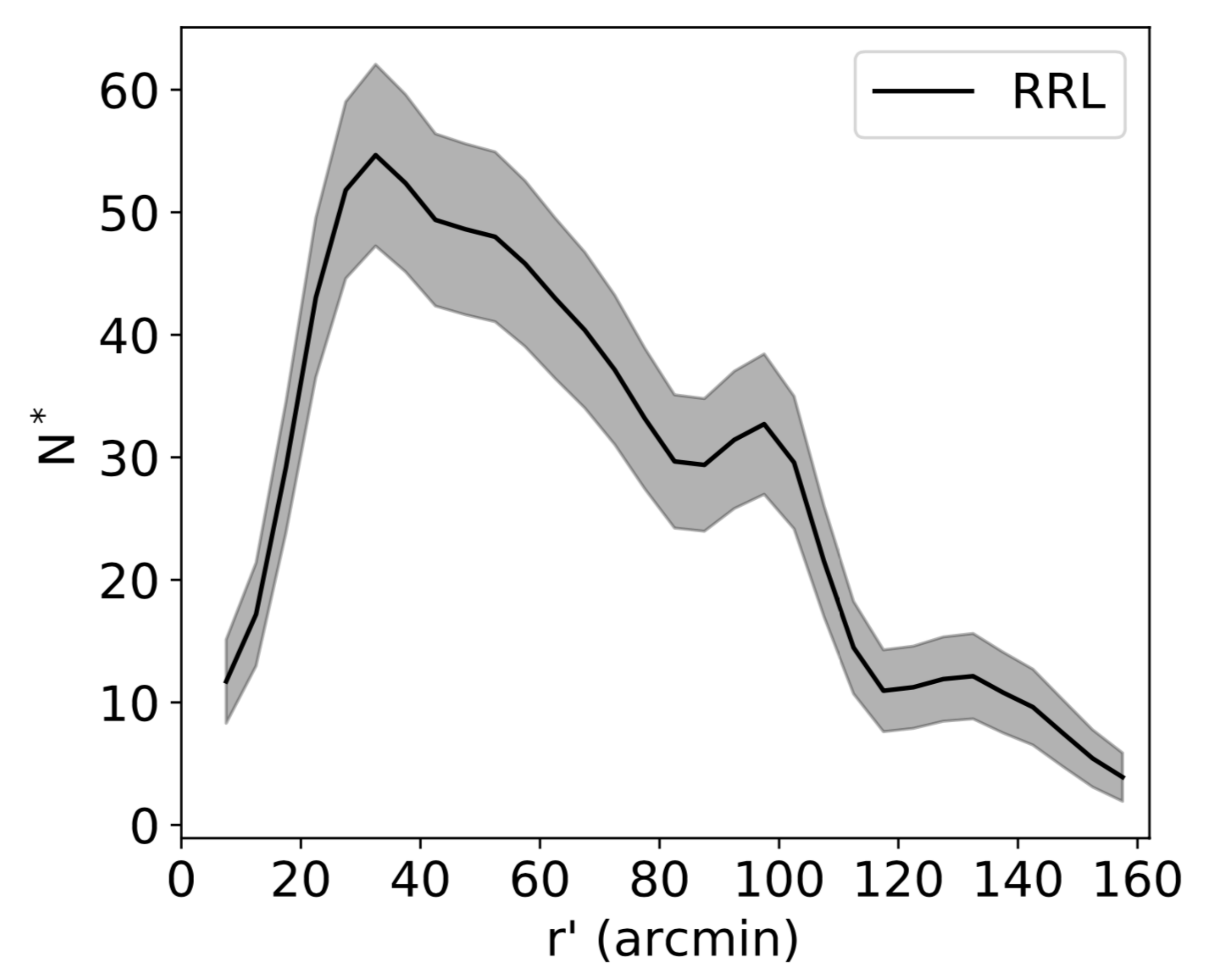}
\caption{Running average of the number of RRL stars as a function of the elliptical distance (r'). The shaded region shows the Poisson uncertainties.}
\label{fig:profile}
\end{figure}

The top panel of Figure~\ref{fig:profile} shows the profile of the number of RRL stars (only \rrab{} and \rrc) as a function of the elliptical distance ($r'$). This distance was obtained taking into account the center, ellipticity and position angle of the galaxy as derived in this work (Table~\ref{tab:ellipse}), and considering it as half the geometrical constant of the ellipse at the location of each individual RRL star. The radial profile was obtained using a running average, sorting the sample in distance, and counting the number of RRL stars within a fixed-size box of $15\arcmin$ and moving in steps of $5\arcmin$. The radial profile appears quite smooth, with just a slight overdensity near $100\arcmin$.

\section{Comparison with a Model of Ant 2's tidal disruption} \label{sec:model}

\begin{figure}
\plotone{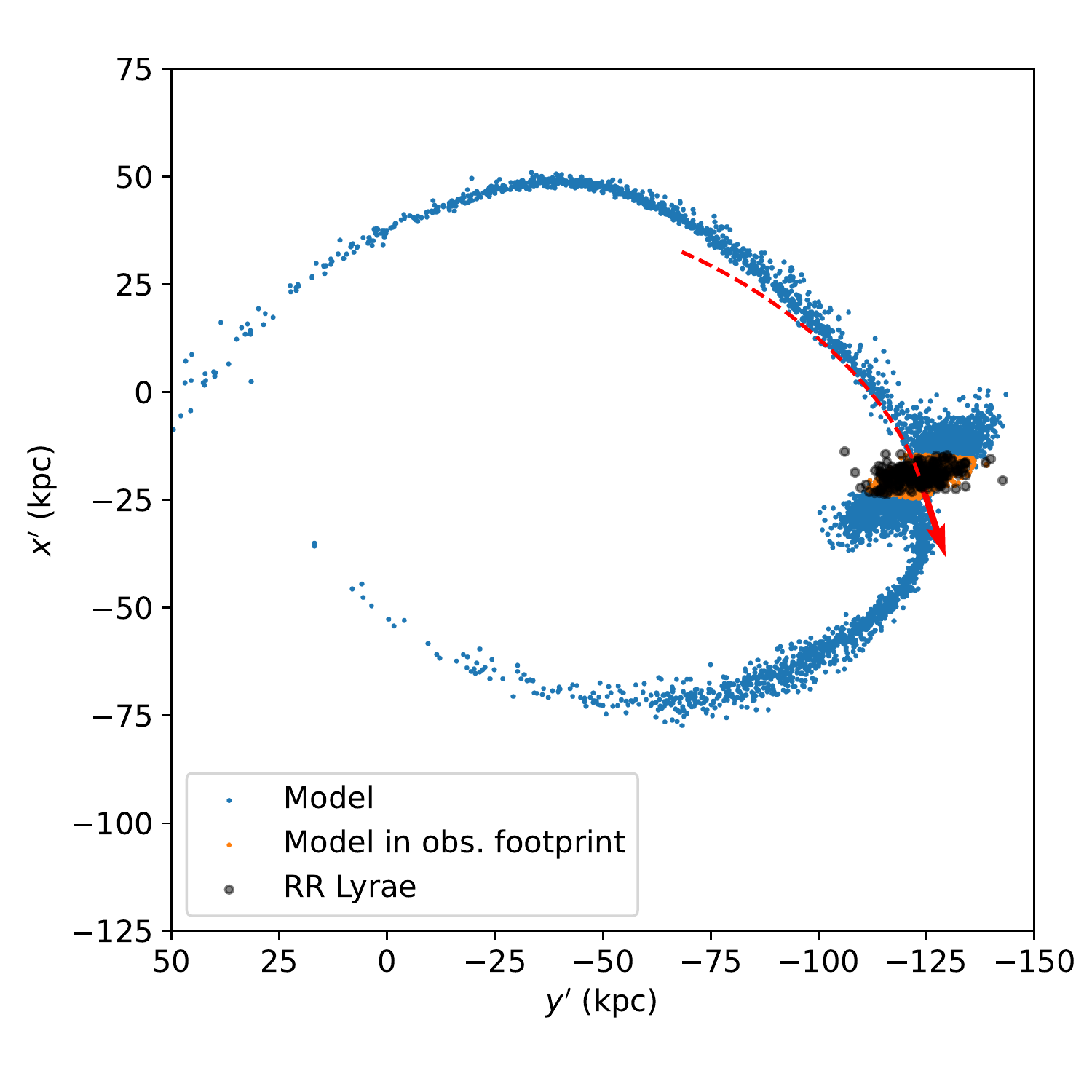}
\caption{Top-down view of Ant 2 from above its orbital plane. The blue points show the stream model while the orange points show the model limited to the sky positions of the RRL stars (black circles). The model matches the observed sample of RRL. The red vector shows the present-day velocity of Ant 2, and the dotted red line shows the past orbit of Ant 2. }
\label{fig:model}
\end{figure}

Next we compare our RRL star sample with models of Ant 2's disruption. We use the same approach as in \cite{Ji21} which simulated the tidal tails of Ant 2 in the presence of the MW and the Large Magellanic Cloud (LMC). These models use the modified Lagrange Cloud Stripping algorithm of \cite{gibbons2014} which was generalized in \cite{erkal2019} to include the effect of the LMC. This algorithm works by integrating the progenitor backwards from the present day location of Ant 2 for 5 Gyr in the presence of the MW and LMC. The system is then evolved forwards in time with particles ejected from the Lagrange points to generate the tidal debris around Ant 2. As in \cite{Ji21}, we model Ant 2 as a Plummer sphere with an initial mass of $10^{7.92} M_\odot$ and a scale radius of 1 kpc. The MW is modelled using the results of \cite{mcmillan2017}. As in \cite{Ji21}, we use a MW model drawn from the posterior chains of \cite{mcmillan2017} which several works have shown is able to fit multiple streams in the MW \citep[e.g.][]{li21,shipp21}\footnote{See Table A3 in \cite{shipp21} for the potential parameters}. This MW has a mass of $8.27\times10^{11} M_\odot$ and gives a realistic past orbit of the LMC, consistent with it being on first approach to the MW \citep[e.g.][]{besla07}. We evaluate the accelerations in this potential using \verb|galpot| \citep{dehnen1998}. We model the LMC as a Hernquist profile with a mass of $1.5\times10^{11} M_\odot$ and a scale radius of 17.14 kpc motivated by the results of \cite{erkal2019}. We place the Sun at a Galactocentric distance of $8.122$ kpc with a velocity of $(11.1,245.04,7.25)$ km s$^{-1}$. For the LMC, we use a present day proper motion, distance, and radial velocity from \cite{Kallivayalil13}, \cite{Pietrzyski13}, and \cite{vanderMarel02} respectively.

For Ant 2, we assume a distance of 124.1 kpc from Table~\ref{tab:distance}, and use proper motions and radial velocities from \cite{Ji21}, $(\mu_\alpha^*,\mu_\delta) = (-0.094,0.103)$ mas yr$^{-1}$, $v_r = 288.8$ km s$^{-1}$ respectively. The resulting tidal debris is shown in Figure \ref{fig:model}, alongside the RRL stars identified in this work. We note that the coordinates shown here are a top-down view of the present-day orbital plane of Ant 2. The red vector shows the present day velocity vector of Ant 2 and the dashed-red curve shows Ant 2's past orbit. The black points show the RRL stars, the blue points show the predicted tidal debris of Ant 2, and the orange points show the predicted debris within the on-sky footprint of the RRL star sample. This shows that within the orbital plane of Ant 2, the RRL stars are extended almost perpendicular to the present day orbit of Ant 2, in agreement with the model. In the model, the cloud of tidal debris close to the progenitor was stripped during the previous pericentric passage, $\sim800$ Myr ago, while the thinner stream was stripped during the penultimate passage, $\sim 3$ Gyr ago. Thus, the observed RRL stars were likely stripped during the most recent pericenter. A future, wider area survey around Ant 2 would be useful to search for the rest of the debris stripped during this pericentric passage and further confirm that Ant 2 is tidally disrupting.

\begin{figure}
\plotone{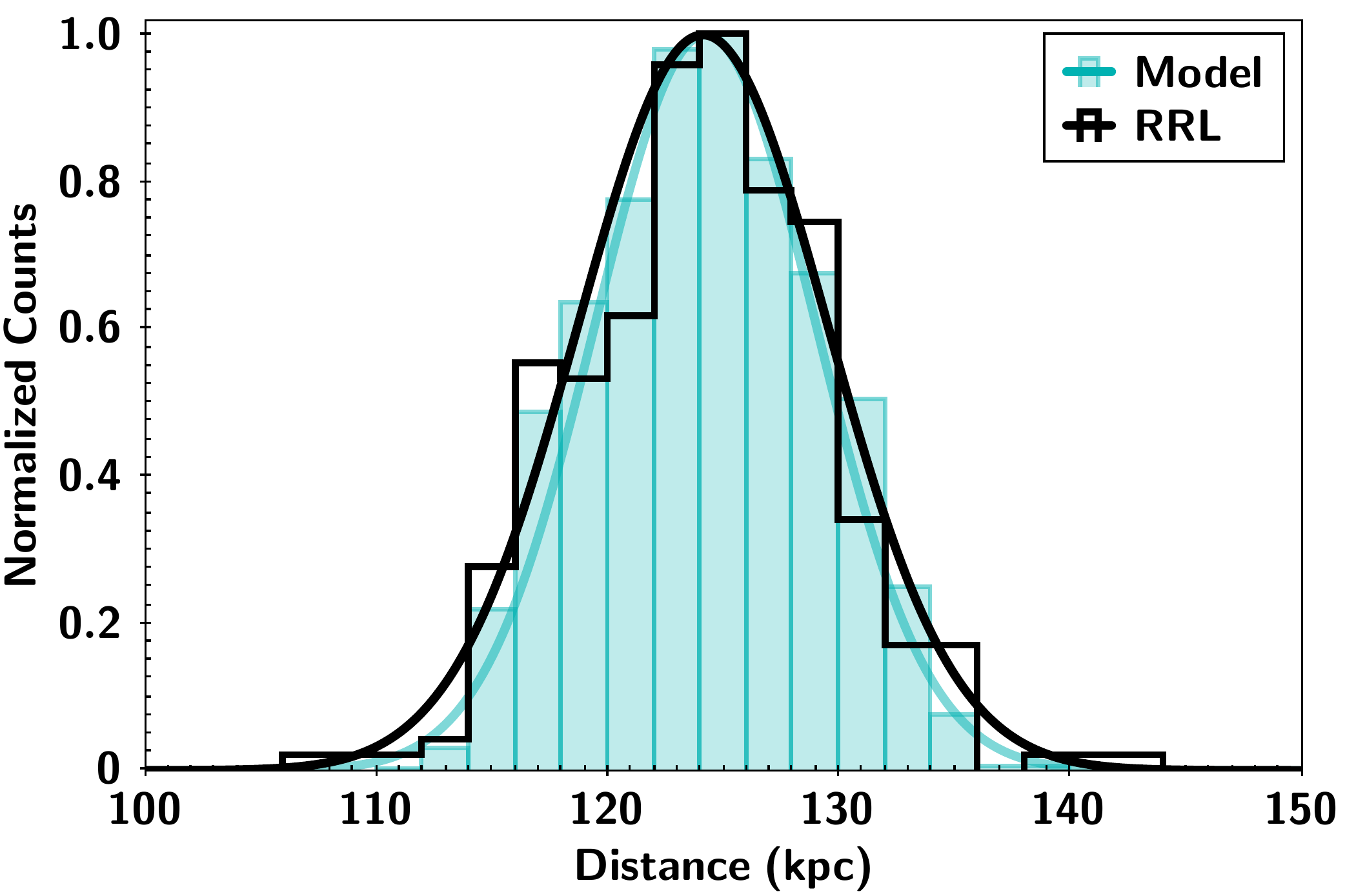}
\caption{Distance distribution of the RRL stars in Ant 2 and the model particles within the same footprint of our survey.}
\label{fig:depth}
\end{figure}

The model predicts quite well the observed depth along the line of sight. In Figure~\ref{fig:depth} we show the histogram of the distribution of distances of the RRL stars in Ant 2, and the one from the model particles in the same footprint of our survey (the orange dots in Figure~\ref{fig:model}). The width of both distributions is strikingly similar.

\section{Anomalous Cepheids in Ant 2} \label{sec:AC}

We have found eight pulsating variable stars in the footprint with magnitudes $r$ between 19.5 and 20.8, corresponding to 0.3-1.6 mags above the HB, which is the expected place for AC stars in a CMD \citep{catelan15}.

There are enough observational similarities between AC and RRL stars that it is worth to discuss with some detail the origin of these stars. Indeed, the first hint that led to the discovery of Ant 2 by \citet{T19} was a group of three RRL stars in the \textit{Gaia} catalog \citep{clementini19} with magnitudes $G\sim 20.2$. A group of three RRL stars, very close together in the sky, with similar magnitudes, and similar proper motions, prompted the search for an underlying stellar population, which was indeed quickly identified.  However, it was clear that the group of RRL stars were much closer than the main body of Ant 2, at distances ranging between 50-95 kpc. The authors speculated that the group of RRL stars may be part of material stripped out from Ant 2.  In this paper, these same three RRL stars are among the group we are re-classifying  as AC members of Ant 2.

\begin{figure}
\plotone{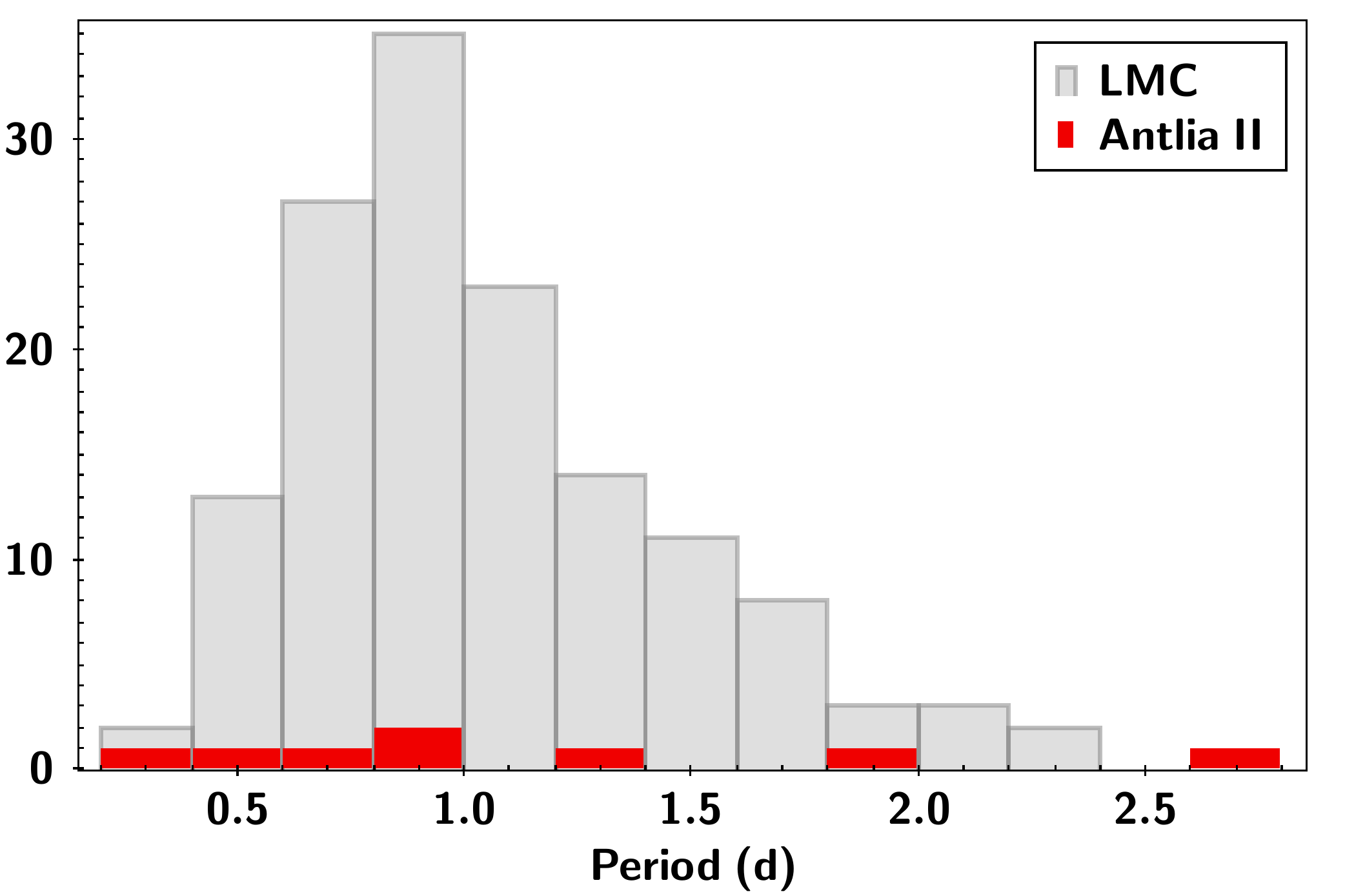}
\caption{Period distribution of AC in the Large Magellanic Cloud \citet{soszynski17a} and Ant 2 (this work.}
\label{fig:ACperiods}
\end{figure}

The main clue that this group of stars are AC and not RRL stars is the period distribution (Figure~\ref{fig:ACperiods}). Although both types of variables have overlapping period distributions, ACs can have periods as long as $\sim 2.5$ d \citep{catelan15}. RRL stars on the other hand have periods $<1.0$ d. In Figure~\ref{fig:ACperiods} we show the period distribution of our sample of AC in Ant 2 compared to that for the LMC from the OGLE survey \citep{soszynski17a}.  The presence of three long-period variables ($>1$ d) in Ant 2 fits well with the expected distribution of periods for AC found in the LMC.

AC are rare stars but they have been found in dSph galaxies around the MW, including the Magellanic Clouds \citep{soszynski17a}, Sculptor \citep{martinez2016b}, Carina \citep{vivas13,coppola15}, Crater 2 \citep{vivas20}, Sextans \citep{mateo95,vivas19}, Draco \citep{kinemuchi08}, and also in satellites of M31 \citep{martinez17}. It is thus not surprising that a galaxy as massive as Ant 2 will also contain a population of these variables. AC have two main formation channels. They may originate as evolved metal-poor stars of an  intermediate-age population \citep[][and references therein]{fiorentino12a}, or alternatively originate from binary evolution of old stars \citep[eg.][]{gautschy17}. There are no obvious hints of an intermediate-age population from the CMD of Ant 2, although the large field contamination may hide such features, particularly if its overall contribution to the galaxy population is small.  With the current data we cannot with certainty discard the intermediate-age formation channel for the AC.

\begin{figure}
\plotone{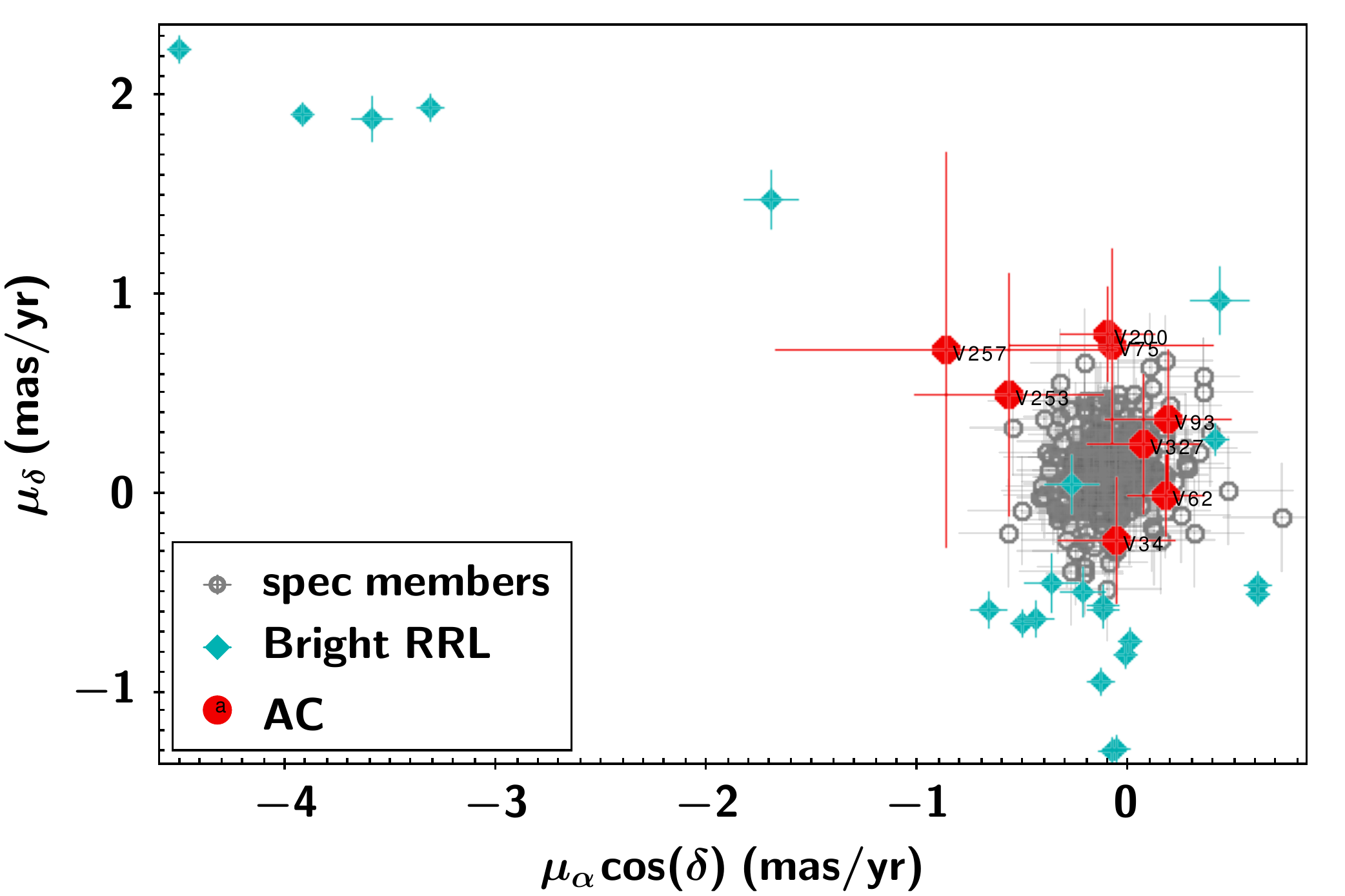}
\caption{Gaia EDR3 proper motions of AC stars in Ant 2 (red circles). For reference the plot also shows the proper motions of spectroscopic members selected by \citet[grey open circles,][]{Ji21}, and other bright RRLs in the field of Ant 2 (cyan diamonds).}
\label{fig:ACpm}
\end{figure}

A final confirmation that the group of AC are members of Ant 2 come from their proper motions. All the stars in this group are sufficiently bright to have proper motion measurements in Gaia EDR3 \citep{gaiaedr3}. In Figure~\ref{fig:ACpm} we show that the proper motions of all eight AC are consistent with those for stars confirmed as radial velocity members of the galaxy \citep{Ji21}, although the errors for the faintest of our AC are quite high. In contrast, the proper motions of other bright RRL stars in the field are more widely distributed. Thus, the proper motions support this group of stars, or most of them, being members of Ant 2 rather than field halo stars. In any case, the latter is an unlikely scenario since halo RRL stars are rare at such large distances. We again estimated the expected number of RRL stars by integrating the number density radial profile of RRL from \citet{medina18} between 50 and 100 kpc, which resulted in $1.4$ RRL stars over the full area of our survey. At most a contamination of two Galactic halo RRL stars is expected among the group of eight AC.

\section{Conclusions} \label{sec:conclusion}

We present a variability survey of 12 sq. deg. around the Ant 2 satellite galaxy, which reveled a large population of RRL stars (318) as well as eight AC. Ant 2 is a very large galaxy, larger than even the SMC, but with a much lower surface brightness. Additional challenges to study this galaxy are its large distance (HB at $r\sim 21.3$) and a position in the sky at low galactic latitude ($b\sim 14 \degr$), which results in a high contamination by field stars. The large FoV of DECam was crucial to be able to survey the full galaxy over just a few days, obtaining time series with excellent S/N at the level of the HB of Ant 2. The RRL stars found in Ant 2 are considered a very pure sample of members since virtually no contamination by MW RRL stars is expected at a similar distance. Based on the RRL stars we derived a distance of $124.1$ kpc, with a dispersion of 5.4 kpc. 

The variable star population of Ant 2 reveals some interesting facts about this galaxy: 

\noindent
{\sl (i)} Ant 2 is large. We find RRL stars over the full surveyed area. The $3\sigma$ elipse that encloses the RRL stars has a major axis over $5\degr$ on the sky. Although there is a decline on the number of RRL stars with radius, it is very likely that there will be more such stars awaiting discovery outside the observed area. 

\noindent 
{\sl (ii)} Ant 2 has a population of AC. First interpreted  by \citet{T19} as foreground RRL stars, we demonstrated here that it is more likely they are AC members of Ant 2. The existence of this population however does not necessarily mean that there is an intermediate-age population in the galaxy, a common interpretation when AC are present, as they also originate from evolution of binary stars in an old population. The current data does not allow us to differentiate with certainty between the formation channels. 

The previous interpretation of these stars being RRL stars was consistent with Ant 2 having lost most of its mass \citep[$\sim90\%$][]{T19}. In such scenario there would be Ant 2 tidal material in front of the main body of the galaxy. New metallicity measurements, and updated proper motions indicate that although the galaxy may be disrupting, such a large mass loss is unlikely \citet{Ji21}, making the existence of debris in front of the galaxy hard to explain. The new interpretation of these stars being AC in Ant 2 solves that controversy since it does not imply there is material in front of the galaxy.

\noindent
{\sl (iii)} Ant 2 is disrupting. The spatial distribution of the RRL stars reveals an elongation ($\epsilon=0.28$), which is approximately aligned with the reflex corrected proper motions of Ant 2 \citep[see][for more details]{Ji21}. In addition, the precise distances of the RRL allow us to detect the elongation along the line of sight which is almost perpendicular to the orbit of Ant 2. Along similar lines, there is a significant gradient in distance along the major axis of the galaxy. Based on an extensive spectroscopic survey covering a similar area to the present work, \citet{Ji21} also detected a gradient along the major axis but for the radial velocity of RGB stars. Both gradients, distance and radial velocity, can be reproduced by a model of the disruption of Ant 2. In this model, the RRL stars we are observing here were torn apart from the main body of Ant 2 during the last pericenter passage, about 800 Myr ago.

The disruption model of Ant 2 predicts long tidal tails coming from the center of the galaxy and extending over more than $100\deg$ in the sky. To search for RRL stars in the tails will be difficult with current instrumentation because of the large search volume. However, the tidal tails of Ant 2, if they indeed exist, should be easily revealed by data from the Vera C. Rubin Observatory after 1-2 years of survey operations.  

\begin{acknowledgments}
This project used data obtained with the Dark Energy Camera (DECam),
which was constructed by the Dark Energy Survey (DES) collaboration.
Funding for the DES Projects has been provided by 
the U.S. Department of Energy, 
the U.S. National Science Foundation, 
the Ministry of Science and Education of Spain, 
the Science and Technology Facilities Council of the United Kingdom, 
the Higher Education Funding Council for England, 
the National Center for Supercomputing Applications at the University of Illinois at Urbana-Champaign, 
the Kavli Institute of Cosmological Physics at the University of Chicago, 
the Center for Cosmology and Astro-Particle Physics at the Ohio State University, 
the Mitchell Institute for Fundamental Physics and Astronomy at Texas A\&M University, 
Financiadora de Estudos e Projetos, Funda{\c c}{\~a}o Carlos Chagas Filho de Amparo {\`a} Pesquisa do Estado do Rio de Janeiro, 
Conselho Nacional de Desenvolvimento Cient{\'i}fico e Tecnol{\'o}gico and the Minist{\'e}rio da Ci{\^e}ncia, Tecnologia e Inovac{\~a}o, 
the Deutsche Forschungsgemeinschaft, 
and the Collaborating Institutions in the Dark Energy Survey. 
The Collaborating Institutions are 
Argonne National Laboratory, 
the University of California at Santa Cruz, 
the University of Cambridge, 
Centro de Investigaciones En{\'e}rgeticas, Medioambientales y Tecnol{\'o}gicas-Madrid, 
the University of Chicago, 
University College London, 
the DES-Brazil Consortium, 
the University of Edinburgh, 
the Eidgen{\"o}ssische Technische Hoch\-schule (ETH) Z{\"u}rich, 
Fermi National Accelerator Laboratory, 
the University of Illinois at Urbana-Champaign, 
the Institut de Ci{\`e}ncies de l'Espai (IEEC/CSIC), 
the Institut de F{\'i}sica d'Altes Energies, 
Lawrence Berkeley National Laboratory, 
the Ludwig-Maximilians Universit{\"a}t M{\"u}nchen and the associated Excellence Cluster Universe, 
the University of Michigan, 
NSF’s NOIRLab, 
the University of Nottingham, 
the Ohio State University, 
the OzDES Membership Consortium
the University of Pennsylvania, 
the University of Portsmouth, 
SLAC National Accelerator Laboratory, 
Stanford University, 
the University of Sussex, 
and Texas A\&M University.

Based on observations at Cerro Tololo Inter-American Observatory, NSF’s NOIRLab (NOIRLab Prop. ID 2018B-0941; PI: A. Walker), which is managed by the Association of Universities for Research in Astronomy (AURA) under a cooperative agreement with the National Science Foundation.
\end{acknowledgments}

\vspace{5mm}
\facilities{Blanco (DECam)}

\software{TOPCAT \citet{taylor05}; PHOTRED \citet{nidever17}}

\appendix 

\section{Time series data} \label{sec:data}

\begin{figure}
\figurenum{16}
\plotone{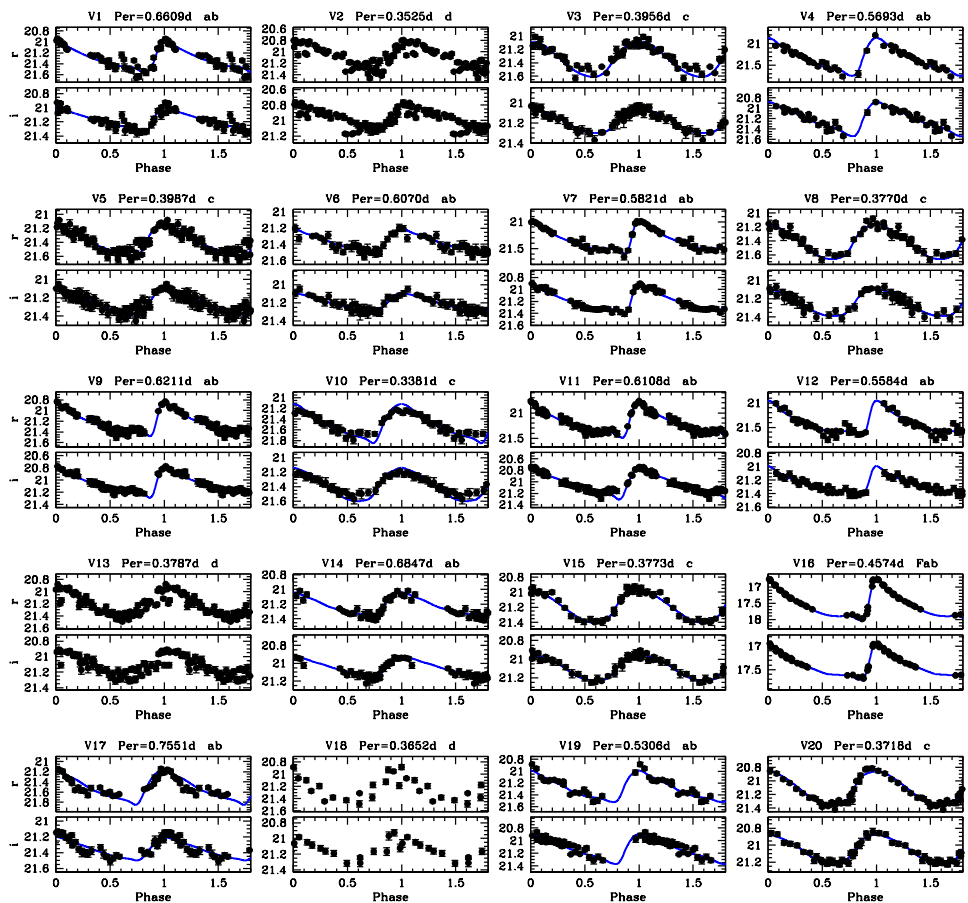}
\label{fig:figureset}
\caption{For each star the $r$, $i$ data are displayed in the top/bottom panels. For all the types $ab$ and $c$ the underlying blue line is the best fitted template to the light curve. The complete figure set (18 images) is available in the online journal.}
\end{figure}

Table~\ref{tab:timeseries} presents the individual epoch photometry in $r$ and $i$ for all the periodic variable stars measured in this work. Light curves for all the stars are provided as Figure Set~\ref{fig:figureset}.

\begin{deluxetable}{rccccccccccccr}
\tabletypesize{\footnotesize}
\tablecolumns{12}
\tablewidth{0pc}
\tablecaption{Time series photometry of variable stars in the field of view of Ant 2.\label{tab:timeseries}}
\tablehead{
ID & Filter & HJD & Mag & Error \\
   &        &  (d) & (mag) & (mag) \\
}
\startdata
  V1 & r &  2458472.709444 &   21.454 &    0.033 \\ 
  V1 & r &  2458472.753670 &   21.087 &    0.016 \\ 
  V1 & r &  2458472.797899 &   20.996 &    0.016 \\ 
  V1 & r &  2458472.842118 &   21.133 &    0.013 \\ 
  V1 & r &  2458504.731142 &   21.265 &    0.016 \\ 
  V1 & r &  2458504.776744 &   21.289 &    0.020 \\ 
  V1 & r &  2458504.821390 &   21.448 &    0.055 \\ 
  V1 & r &  2458504.865625 &   21.499 &    0.047 \\ 
\enddata
\tablecomments{Table~\ref{tab:timeseries} is published in its entirety in the machine-readable format. A portion is shown here for guidance regarding its form and content.}
\end{deluxetable} 

\end{document}